\documentclass[conference,compsoc]{IEEEtran}

\ifCLASSOPTIONcompsoc
  \usepackage[nocompress]{cite}
\else
  \usepackage{cite}
\fi
\usepackage{amssymb,amsfonts}
\usepackage{algorithmic}
\usepackage{algorithm}
\usepackage{subfigure}
\usepackage{threeparttable}
\usepackage{multirow}
\usepackage{bm}
\usepackage{graphicx}    
\usepackage{pifont}
\usepackage{float}
\usepackage{soul, color,xcolor}
\usepackage{tabularx}
\usepackage{enumitem}
\usepackage{listings}
\usepackage{subcaption}
\usepackage{filecontents}
\usepackage{booktabs}
\usepackage{amsmath}
\usepackage{picins}
\usepackage{framed}

\makeatletter
\newcommand{\rmnum}[1]{\romannumeral #1}

\newcommand{\Rmnum}[1]{\expandafter\@slowromancap\romannumeral #1@}
\makeatother

\ifCLASSINFOpdf
\else
\fi

\begin{document}

\title{Chameleon: An Efficient FHE Scheme Switching Acceleration on GPUs}

\author{\IEEEauthorblockN{Zhiwei Wang\IEEEauthorrefmark{2},
Haoqi He\IEEEauthorrefmark{2},
Lutan Zhao\IEEEauthorrefmark{2},
Peinan Li\IEEEauthorrefmark{2},
Zhihao Li\IEEEauthorrefmark{4},
Dan Meng\IEEEauthorrefmark{2}, 
Rui Hou\IEEEauthorrefmark{2}}
\IEEEauthorblockA{\IEEEauthorrefmark{2}Key Laboratory of Cyberspace Security Defense, Institute of Information Engineering, CAS\\
and University of Chinese Academy of Sciences;
\IEEEauthorrefmark{4}Ant Group.}
\{wangzhiwei, hehaoqi, zhaolutan, lipeinan, mengdan,  hourui\}@iie.ac.cn, lzh458070@antgroup.com
}

\maketitle

\begin{abstract}
Fully homomorphic encryption (FHE) enables direct computation on encrypted data, making it a crucial technology for privacy protection. 
However, FHE suffers from significant performance bottlenecks. 
In this context, GPU acceleration offers a promising solution to bridge the performance gap. 
Existing efforts primarily focus on single-class FHE schemes, which fail to meet the diverse requirements of data types and functions, prompting the development of hybrid multi-class FHE schemes.
However, studies have yet to thoroughly investigate specific GPU optimizations for hybrid FHE schemes.

In this paper, we present an efficient GPU-based FHE scheme switching acceleration named Chameleon.
First, we propose a scalable NTT acceleration design that adapts to larger CKKS polynomials and smaller TFHE polynomials. 
Specifically, Chameleon tackles synchronization issues by fusing stages to reduce synchronization, employing polynomial coefficient shuffling to minimize synchronization scale, and utilizing an SM-aware combination strategy to identify the optimal switching point.
Second, Chameleon is the first to comprehensively analyze and optimize critical switching operations. 
It introduces CMux-level parallelization to accelerate LUT evaluation and a homomorphic rotation-free matrix-vector multiplication to improve repacking efficiency.
Finally, Chameleon outperforms the state-of-the-art GPU implementations by 1.23$\times$ in CKKS HMUL and 1.15$\times$ in bootstrapping. 
It also achieves up to 4.87$\times$ and 1.51$\times$ speedups for TFHE gate bootstrapping compared to CPU and GPU versions, respectively, and delivers a 67.3$\times$ average speedup for scheme switching over CPU-based implementation.

\end{abstract}

\IEEEpeerreviewmaketitle

\vspace{-0.15cm}
\section{Introduction}
\vspace{-0.15cm}

Fully homomorphic encryption (FHE) supports direct computation on encrypted data, making it a highly regarded technology for data privacy protection \cite{gentry2009fully}. 
However, it encounters significant performance bottlenecks that hinders its widespread adoption in practical applications.
To address this issue, researchers attempt to utilize heterogeneous platforms (e.g., FPGA\cite{riazi2020heax,agrawal2022fab,yang2023poseidon}, ASIC\cite{samardzic2021f1,kim2022bts,kim2023sharp}, and GPU\cite{wang2023he,al2018high,fan2023tensorfhe,morshed2020cpu}) to accelerate FHE. 
Among them, GPUs are preferred candidates due to their flexible algorithm adaptability and the availability of off-the-shelf products.

Existing efforts primarily focus on accelerating individual FHE schemes, either word-wise schemes that support SIMD linear operations (e.g., BGV\cite{wang2023he}, BFV\cite{al2018high}, and CKKS\cite{fan2023tensorfhe}), or bit-wise schemes that accommodate nonlinear operations (e.g., TFHE\cite{morshed2020cpu}).
However, single-class FHE schemes pose challenges in effectively meeting the diverse requirements of real-world applications with data types and functional capabilities\cite{lee2022low,cheon2023high,lee2023precise,bourse2018fast,lou2019she}.

To address these limitations, researchers propose a series of feasible scheme switching algorithms that combine different types of FHE schemes~\cite{boura2020chimera,lu2021pegasus,chen2021efficient,kim2023general}. 
{However, current implementations remain inefficient due to the inherent performance barriers of FHE.
Even with GPUs, it is essential to carefully consider the significant differences in ciphertext sizes and homomorphic operations among various FHE schemes. 
In addition, critical switching paths should be thoroughly examined to identify potential bottlenecks.}
Below, we highlight two critical problems.

\textbf{Problem 1: Single NTT acceleration struggles to achieve optimal performance across different ciphertext sizes.} 
NTT is the most performance-critical operator for FHE, contributing over 70\% execution time for bootstrapping~\cite{kim2022ark}.
{However, traditional GPU-based implementations face significant synchronization issues, such as excessive synchronizations (e.g., up to ${\rm log}_2N-1$ synchronizations).}
Stall overhead from thread synchronization accounts for 35\% of the total time~\cite{wang2023nttfusion}.
Moreover, polynomial lengths vary considerably, with CKKS generally employing larger \(2^{16}\) and TFHE utilizing smaller \(2^{10}\).
Thus, a singular design must enhance its adaptability to perform optimally in all situations, posing challenges for developing further solutions.

\textbf{Problem 2: Critical switching operations fail to exploit data-level parallelism while also neglecting algorithmic optimizations.}
The traditional slot-level approach for LUT evaluation processes each TFHE ciphertext individually, requiring frequent kernel launches and repeated access to bootstrapping keys. 
This method fails to fully leverage the inherent parallelism of independent TFHE ciphertexts.
In addition, the existing linear transformation (the core part of repack) utilizes a matrix-vector multiplication based on the BSGS algorithm. 
This approach still necessitates numerous costly homomorphic rotations, leading to inefficient conversion from CKKS ciphertexts to TFHE ciphertexts.

{Focusing on the above issues, we can leverage two key observations: the stage-structured fusion properties of NTT and ciphertext independence in critical switching operations. }
\textbf{Observation 1:} 
While fusing multiple stages appears to reduce synchronization, it does not always yield optimal results.
For instance, parallelism should be prioritized for small TFHE polynomials, while avoiding thread divergence becomes crucial for large CKKS polynomials.
In addition, we observe that restructuring the data layout can reduce the inter-thread synchronization scale across different thread blocks. 
Besides, investigating the distribution of synchronization points offers opportunities to capture the optimal switching points.
\textbf{Observation 2:} 
During LUT evaluation, independent TFHE ciphertexts share the same gate bootstrapping and repeatedly access the bootstrapping keys for internal CMux gates. 
This insight allows multiple CMux gates to be merged into a single gate, thereby facilitating the batch processing of ciphertexts.
For repack, the advantageous property of automorph can be leveraged to precompute all rotation ciphertexts of the LWE secret key used in the linear transformation, thereby reducing the number of homomorphic rotations.

Based on these observations, we propose the first efficient GPU-based FHE scheme switching acceleration named {Chameleon}. 
{It introduces tailored NTT synchronization optimizations that adapt to polynomial transform of various sizes in CKKS and TFHE. 
Moreover, Chameleon leverages the parallel potential of critical switching operations (e.g., LUT evaluation and repack) to achieve faster ciphertext conversion.}
The major contributions are as follows:

\begin{itemize}[leftmargin=0.5cm, noitemsep, nolistsep]
    \item 
    \textbf{Propose a scalable NTT acceleration design.}
    First, we introduce tailored stage fusion methods to reduce the number of synchronizations: butterfly decomposition for smaller TFHE polynomials and thread aggregation for larger CKKS polynomials.
    Second, a polynomial coefficient shuffling method is utilized to restructure the data layout and minimize the synchronization scale.
    Third, we present an SM-aware synchronization combination strategy that explores possible combinations to identify the optimal switching point, defined as the number of split data chunks closest to the number of SMs. 
    
    \item 
    \textbf{Propose LUT evaluation with CMux gate-level parallelism.}
    Conventional slot-level methods process each TFHE ciphertext individually, leading to repeated execution of core operators and redundant access to bootstrapping keys during TFHE gate bootstrapping. 
    However, we observe that despite having different data, the ciphertexts involved in LUT evaluation share the same CMux operations. 
    Building on this insight, we introduce a novel CMux gate-level parallelization method.
    This approach constructs batched CMux gates to simultaneously process multiple independent TFHE ciphertexts, fully leveraging data-level parallelism.
    
    \item 
    \textbf{Propose repack acceleration with homomorphic rotation-free MatVec optimization.}
    Current BSGS-based MatVec implementation fails to fully exploit the space-time trade-off, as it still results in real-time homomorphic rotation operations.
    We observe that automorph inherently support linear operations, enabling the linear operations in the \textit{Baby} step to be merged with automorphisms in the \textit{Giant} step. Consequently, all rotation ciphertexts corresponding to the combined automorph mapping can be precomputed, eliminating costly homomorphic rotations and reducing the process to inexpensive scalar multiplications.

    \item 
    \textbf{Evaluate performance on a realistic NVIDIA GPU server.}
    Chameleon outperforms state-of-the-art GPU implementations with a 1.23$\times$ speedup in CKKS homomorphic multiplication and a 1.15$\times$ speedup in bootstrapping. 
    It also achieves up to 4.87$\times$ and 1.51$\times$ performance boosts over CPU and GPU versions for TFHE gate bootstrapping and provides a remarkable 67.3× speedup for scheme switching compared to CPU-based implementations.
\end{itemize}

\vspace{-0.15cm}
\section{Background}
\vspace{-0.15cm}

In this section, we briefly review the CKKS and TFHE schemes and scheme switching algorithms.

\vspace{-0.15cm}
\subsection{CKKS Scheme}
\vspace{-0.15cm}

{\textbf{Ciphertext structure.} 
CKKS can pack \(n_{\rm slot}\) real or complex numbers into a vector, referred to as the message \(m\). 
During the encryption, the message \(m\) is encoded into an integer polynomial \(P(x)\) of degree \(n_{\rm ckks}\), where \(n_{\rm ckks}\) is a power of 2 and \(n_{\rm slot} \leq {n_{\rm ckks}}/{2}\).
Note that the polynomial \(P(x)\) is multiplied by a scaling factor \( \Delta \) and then rounded. 
 \( P \) is then encrypted as an RLWE ciphertext, represented by a pair of polynomials \((B(x), A(x)) \in R_Q^2\).
In this case, \( B(x) = A(x) \cdot S(x) + \Delta \cdot P(x) + E(x) \), with the polynomial coefficients reduced modulo the ciphertext modulus \( Q \).}

{\textbf{Basic CKKS primitives.} CKKS supports a range of homomorphic operations, including homomorphic addition ({\texttt{HADD}}) and multiplication ({\texttt{HMUL}}) between ciphertexts, plaintext-ciphertext addition ({\texttt{PADD}}) and multiplication ({\texttt{PMUL}}), as well as homomorphic rotation ({\texttt{HROT}}). 
Note that during {HMult} and {PMult} operations, the internal scaling factor \(\Delta\) is also involved in the computation, resulting in \(\Delta^2\). 
At this point, the \texttt{{Rescale}} operation is required to reduce it back to \(\Delta\).
While this operation maintains the stability of the scaling factor during computation, it also reduces the ciphertext modulus, which decreases its level. 
When only one level remains, a \texttt{{Bootstrapping}} operation is required to restore the level for further computations.}

{\textbf{CKKS Bootstrapping.} 
Generally, it consists of four steps: {{Modulus Raising}}, {{Coeff2Slot}}, {{Approximated modulo}}, and {{Slot2Coeff}}\cite{cheon2018bootstrapping}.
This operation raises the ciphertext modulus to a sufficiently high level, allowing for more homomorphic operations and supporting the multiplication depth required in complex applications. 
However, bootstrapping remains computationally expensive due to the resource-intensive homomorphic operations involved, particularly multiplications and rotations.}

\vspace{-0.2cm}
\subsection{TFHE Scheme}
\vspace{-0.2cm}

\textbf{Ciphertext structure.} 
{TFHE typically includes three types of ciphertexts:}
\ding{172} {\textit{LWE ciphertext.} 
Unlike the polynomial representation of RLWE ciphertexts, LWE ciphertexts are expressed as \((a_0, a_1, a_2, \cdots, a_{n_{\rm lwe}-1}, b) \in \mathbb{Z}_q^{n_{\rm lwe} + 1}\), where \(b = \sum_{i=0}^{n_{\rm lwe}}a_i \cdot s_i + m + e\). 
Here, \(m\) represents the scalar message, \(e\) is the error term, and \(n_{\rm lwe} + 1\) denotes the dimension.}
\ding{173} {\textit{RLWE ciphertext.}
Similar to CKKS, TFHE RLWE ciphertext is also represented as a polynomial pair \((B(x), A(x))\) with a dimension of $N$.}
\ding{174} {\textit{RGSW ciphertext.}
RGSW ciphertexts can be viewed as a two-dimension matrix of RLWE ciphertexts, typically sized \((k + 1) \times (k + 1) \cdot l\), where \(k = 1\) and \(l\) represents the decomposition level. 
In TFHE, bootstrapping keys are commonly represented as RGSW ciphertexts.}

{\textbf{TFHE bootstrapping.}
It is usually performed immediately after a boolean operation on LWE ciphertext to reduce noise in the resulting ciphertext. 
Unlike CKKS bootstrapping, TFHE gate bootstrapping involves more fine-grained operations. 
Specifically, it involves the following steps:
\ding{172} \texttt{{Modulus Switching}}:  
Each element of the LWE ciphertext is converted from modulus \(q\) to \(2N\) by multiplying by the scaling factor \(2N/q\) and rounding to the nearest integer, which is straightforward as \(N\) is typically a power of 2.
\ding{173} \texttt{Blind Rotation}: 
The scaled LWE ciphertext is first converted into an RLWE ciphertext, which is then combined with the TFHE bootstrapping key in RGSW form to perform the CMux gate operation. 
This process involves an \texttt{{external product}} product between the RLWE ciphertext and the RGSW key. 
The resulting RLWE ciphertext is used as input for the next CMux gate, and this process is repeated \(n_{\rm lwe}\) times.
\ding{174} {\texttt{Sample Extract}}: 
This process extracts constant terms from RLWE ciphertext to create a new LWE ciphertext and then performs a \texttt{key switching} for format compatibility.}

\vspace{-0.25cm}
\subsection{Scheme Switching Algorithm}
\vspace{-0.25cm}

{To harness the strengths of both FHE schemes, researchers devise a range of scheme switching algorithms that facilitate seamless integration of these approaches \cite{boura2020chimera,lu2021pegasus,chen2021efficient}. 
Below, we provide a detailed analysis of these switching algorithms, focusing on CKKS and TFHE.}

{\textbf{Scheme Conversion from CKKS to TFHE.}
CKKS RLWE ciphertext can be converted into multiple TFHE LWE ciphertexts using a \texttt{sample extract} procedure, with the number of LWE ciphertexts corresponding to the number of CKKS slots. 
These LWE ciphertexts are mutually independent, allowing for parallel processing. 
Notably, the slot-encoded CKKS ciphertext must undergo a \texttt{Slot2Coeff} operation before sample extraction, which converts the ciphertext into a coefficient-encoded format.
}

{\textbf{Scheme Conversion from TFHE to CKKS.}
Each extracted LWE ciphertext can be subjected to nonlinear operations through functional bootstrapping by evaluating a \texttt{look-up table(LUT)}. 
In this process, the TFHE bootstrapping procedure incorporates a {blind rotation} step, generating multiple RLWE ciphertexts. 
The extracted TFHE LWE ciphertexts are merged into a single CKKS RLWE ciphertext using a \texttt{repack} operation, mainly involving linear transformations.}

\vspace{-0.15cm}
\subsection{Number Theoretic Transform}
\label{subsec:ntt}
\vspace{-0.15cm}

\textbf{NTT vs. FFT.}
{CKKS and TFHE often require computationally intensive polynomial multiplications within the ring \(R_q = \mathbb{Z}_q[x]/(x^N+1)\), represented as \(c(x) = a(x) \cdot b(x) \mod (x^N+1)\). To compute $c(x)$, FHE accelerators commonly use the Fast Fourier Transform (FFT) or Number Theoretic Transform (NTT) to improve efficiency. These techniques reduce the time complexity of polynomial multiplication from \({O}(N^2)\) to \({O}(N \log N)\). While FFT operates in the complex domain, NTT works in finite fields. Rounding errors in FFT can cause decryption failures, so NTT is generally preferred for CKKS and TFHE schemes.}

\begin{figure*}[t]
\setlength{\belowcaptionskip}{-0.6cm}
\setlength{\abovecaptionskip}{0.1cm}
\centering
\includegraphics[width=0.8\linewidth]{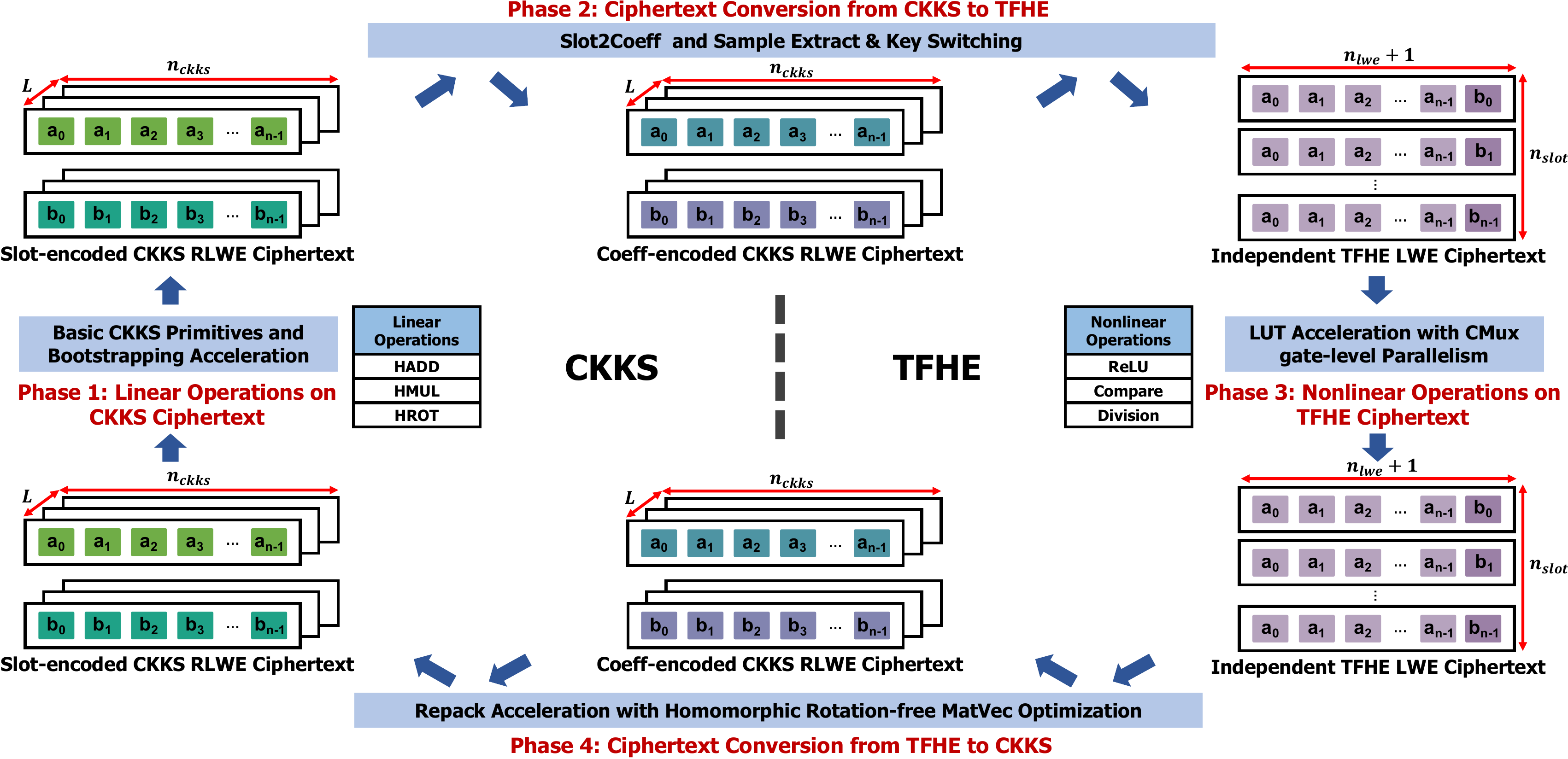}
\caption{{Working flow of {Chameleon}.}} 
\label{fig:overview}
\end{figure*}

\textbf{Negacyclic convolution-based NTT.}
Specifically, Roy et al. \cite{roy2014compact} first present the well-known decimation-in-time (DIT) NTT based on the Cooley-Tukey (CT) butterfly, which takes the input in the natural order and produces the output in bit-reversal order ($no \rightarrow bo$). 
Then, Poppelmann et al. \cite{poppelmann2015high} exploit a decimation-in-frequency (DIF) NTT based on the Gentleman-Sande (GS) butterfly, which takes the input in bit-reversal order and produces the output in natural order ($bo \rightarrow no$). 
Equation \ref{equa:ncc-conv} demonstrates that combining the two implementations achieves polynomial multiplication.
\begin{equation}
    c =  {\rm INTT}_{bo\rightarrow no}^{GS,\psi^{-1}}({\rm NTT}_{no\rightarrow bo}^{CT,\psi}(a) \odot {\rm NTT}_{no\rightarrow bo}^{CT,\psi}(b))
    \label{equa:ncc-conv}
\end{equation}
where the constant $\psi$ is called the twiddle factor, i.e., the primitive $2N$-th root of unity in $\mathbb Z_q$ satisfying the condition $\psi^{2N} = (1\ {\rm mod}\ q)$ and $\psi^i \neq (1\ {\rm mod}\ q)$ $\forall$ $i<2N$.

\vspace{-0.15cm}
\subsection{GPU Overview}
\label{sec:gpu}
\vspace{-0.15cm}

\textbf{Basic GPU architecture.} 
In a typical GPU, a kernel launches thousands of threads for high throughput. 
These threads are grouped into thread blocks (TBs), assigned to Streaming Multiprocessors (SMs). 
While multiple TBs can share the same SM, each TB operates within a single SM. 
Threads within a TB are organized into warps, usually of size 32, and are mapped to Streaming Processors (SPs). 
The SM's warp scheduler determines the execution order of warps to maximize performance.
GPUs also offer a memory hierarchy with different types. 
Frequently accessed data can be stored in shared memory (SMEM) within a thread block for faster access than global memory (GMEM), holding a small NTT sequence. 
The \_\_syncthreads() function provides intra-block synchronization. 
Constant memory (CMEM) is globally accessible, cached, and faster than GMEM. However, it has a 64 KB limit, making it suitable for small constants like twiddle factors.
For more details, refer to \cite{sanders2010cuda}.

{\textbf{Why are tailored GPU optimizations for CKKS and TFHE needed?} 
Due to the differing parameter scales of CKKS and TFHE schemes, their homomorphic operations impose varying hardware resource requirements when mapped onto GPU architectures.
For example, the word-wise CKKS scheme requires large ciphertext moduli (e.g., 2305-bit) and longer polynomials (e.g., $2^{16}$) to ensure security and support bootstrapping \cite{jung2021over}. 
In contrast, the bit-wise TFHE scheme uses smaller moduli (e.g., 32/64 bits) because it encrypts individual bits and performs bootstrapping after each operation, allowing for shorter polynomials, typically around 1024.
Given the substantial differences in polynomial lengths between the two schemes, it is crucial to analyze and develop tailored optimization methods. These efforts will enable dynamic support for scheme switching acceleration.}

\section{Design Overview}
\vspace{-0.3cm}
To meet the diverse requirements for data and function types in practical applications (e.g., private inference~\cite{lee2022low}), combining FHE schemes with varying algorithmic characteristics emerges as an effective solution~\cite{boura2020chimera,lu2021pegasus,chen2021efficient,kim2023general}. 
However, these works exhibit poor ciphertext conversion efficiency due to CPU platforms' limited parallel capabilities. 
To further enhance performance, this paper introduces the first GPU-based FHE scheme switching acceleration named Chameleon.
Figure~\ref{fig:overview} shows our design overview, which consists of four phases, including:

\textbf{Phase 1: Linear operations on CKKS ciphertext.}
It can execute a series of linear operations (e.g., addition, multiplication, and rotation) on the slot-encoded CKKS RLWE ciphertext in a SIMD manner.
Like previous designs, we reconstruct fine-grained operators for CKKS homomorphic operations, mainly focusing on critical key switching. 
This includes components such as \texttt{NTT}, \texttt{BConv}, \texttt{ModMul}, \texttt{ModAdd}, and \texttt{Automorph}. 
Notably, NTT serves as the most beneficial component. 
However, when mapping large polynomials to GPU hardware, the stall overhead from thread synchronization seriously limits performance. 
To solve this problem, we propose a stage fusion-based NTT acceleration design tailored for larger CKKS polynomials. 

\textbf{Phase 2: Ciphertext conversion from CKKS to TFHE.}
The slot-encoded CKKS ciphertext, following linear operations, must undergo conversion into coefficient-encoded CKKS ciphertext via the {Slot2Coeff} operation. 
Subsequently, sample extract and key switching are executed to produce independent TFHE ciphertexts corresponding to the number of slots.

\textbf{Phase 3: Nonlinear operations on TFHE ciphertext.}
Employing traditional slot-level parallel methods for LUT evaluation necessitates the individual mapping of gate bootstrapping for each ciphertext to the GPU. 
However, numerous low-workload TFHE operators (e.g., \texttt{Decompose}, \texttt{NTT}, and \texttt{MAC}) lead to low hardware utilization, elevated launch overhead, and data access redundancy.
To address these issues, we first optimize the NTT operator specifically for smaller TFHE polynomials, focusing on ensuring sufficient thread-level parallelism.
Second, despite differing data, these ciphertexts actually share the same gate bootstrapping operations and keys. 
Leveraging this insight, we propose a novel CMux-level parallel method that constructs a unified CMux gate to batch process all TFHE ciphertexts, effectively harnessing data-level parallelism.

\textbf{Phase 4: Ciphertext conversion from TFHE to CKKS.}
Following the nonlinear operations, multiple independent TFHE LWE ciphertexts can be converted back into slot-encoded CKKS RLWE ciphertexts through the repack operation. 
This process primarily involves matrix-vector multiplication between the plaintext matrix from the LWE ciphertext and the rotation ciphertext of the LWE secret key. 
However, existing BSGS-based implementation remains limited by the high costs of homomorphic rotations. 
By utilizing the inherent properties of automorph that support linear operations, the automorph from the \textit{Giant} step can be combined with linear operations from the \textit{Baby} step.  
Importantly, this allows us to precompute all rotation ciphertext corresponding to the combined automorph mapping.
As a result, we propose a homomorphic rotation-free MatVec implementation to accelerate repack, which only involves inexpensive scalar multiplications.

\vspace{-0.15cm}
\section{Critical Switching Operations Acceleration on GPUs}
\vspace{-0.15cm}

\begin{figure}[t]
\setlength{\belowcaptionskip}{-0.3cm}
\setlength{\abovecaptionskip}{0.1cm}
\centering
\includegraphics[width=\linewidth]{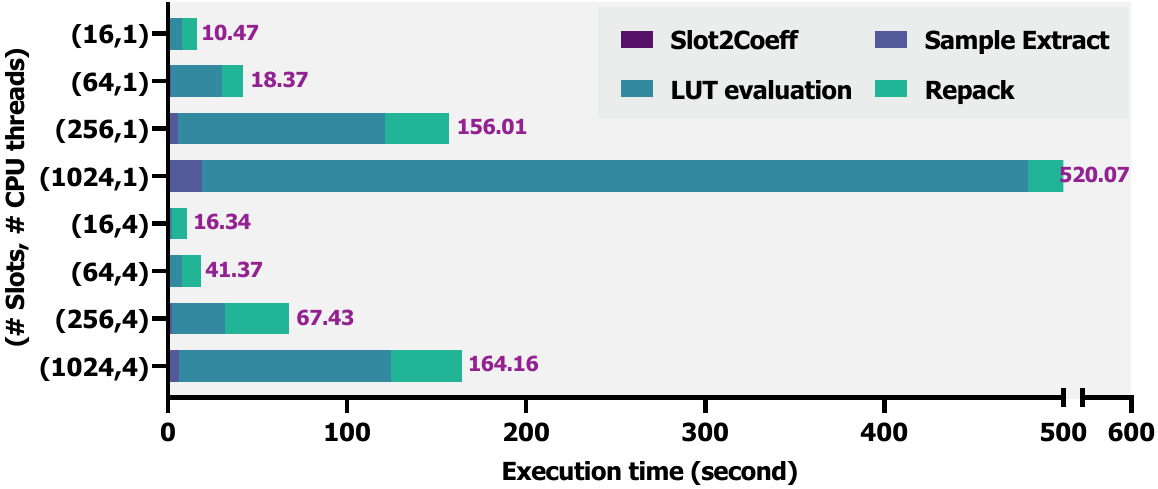}
\caption{Performance breakdown of scheme switching between CKKS and TFHE in {Pegasus} ($n_{\rm ckks}$=$2^{16}$, $n_{\rm lwe}$=$2^{10}$, $n_{\rm lut}$=$2^{12}$, ciphertext modulus=599 bits).}
\label{fig:peg-per}
\end{figure}

We conduct a performance breakdown of scheme switching between CKKS and TFHE in Pegasus~\cite{lu2021pegasus}, as illustrated in Figure~\ref{fig:peg-per}, which reveals three key observations:
\textit{(\rmnum{1}) LUT evaluation and repack operation dominate the entire process, serving as the primary performance bottleneck.} 
Notably, these two operations account for about 96\% of the total time, regardless of the number of slots or threads. 
Therefore, accelerating them is crucial for enhancing overall efficiency.
\textit{(\rmnum{2}) LUT evaluation exhibits rewarding potential for parallel acceleration.}
For instance, increasing the number of slots from 256 to 1024 raises the execution time from 74.1\% to 88.8\%. 
However, with four CPU threads, these times decrease to 43.9\% and 72.2\%, respectively.
\textit{(\rmnum{3}) Repack performs well with fewer slots and remains unaffected by the increased CPU threads, highlighting the necessity for algorithmic optimization.}
At 64 slots, the execution time of repack surpasses that of LUT evaluation, becoming a new performance bottleneck, while its efficiency remains stable with one and four CPU threads.
Based on the above analysis, we propose LUT evaluation with CMux gate-level parallelism and repack with homomorphic rotation-free BSGS optimization. 
The specific details are as follows.

\begin{figure*}[t]
\setlength{\belowcaptionskip}{-0.5cm}
\centering
\includegraphics[width=\linewidth]{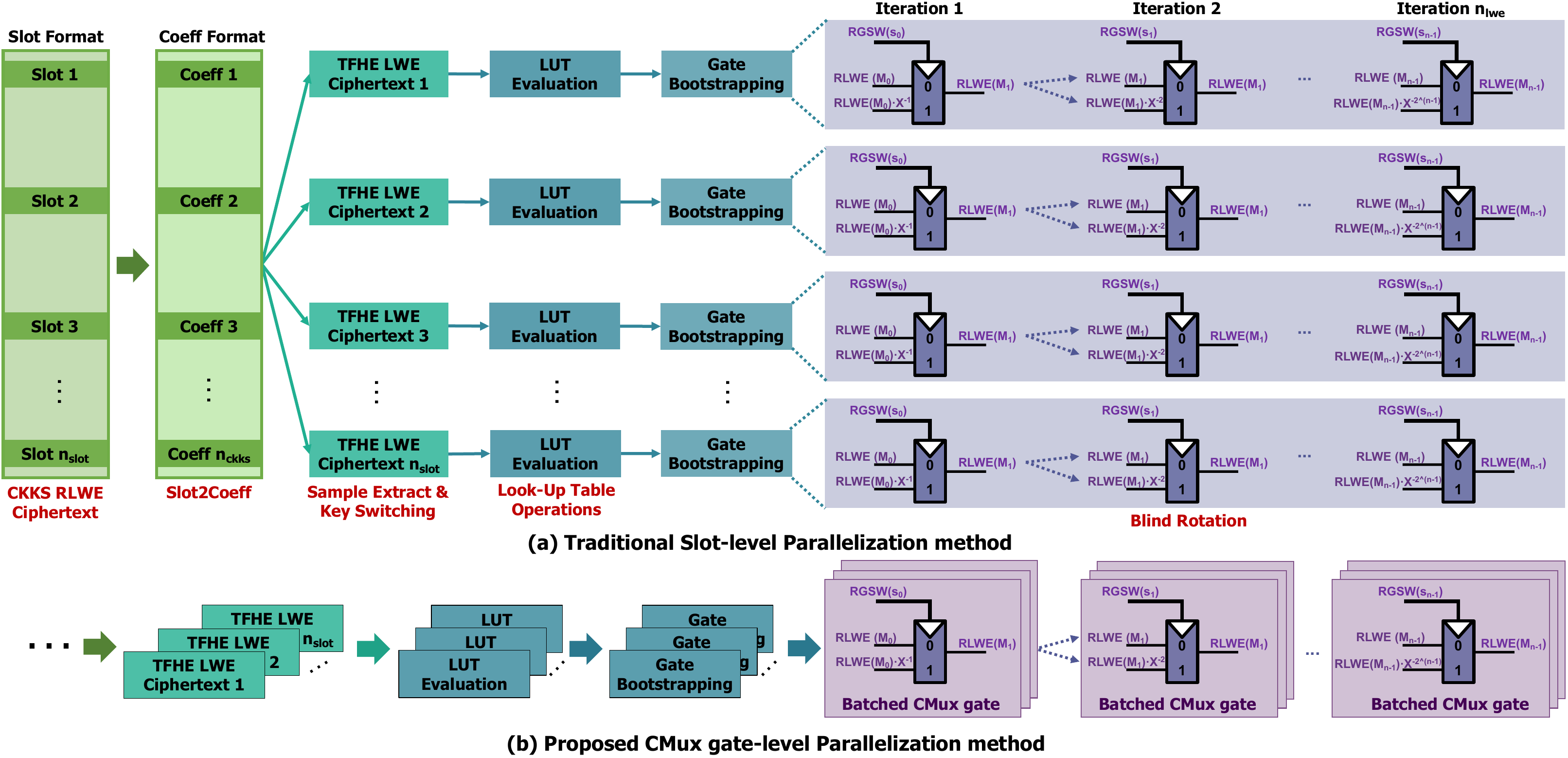}
\caption{LUT acceleration with two levels of parallelism.}
\label{fig:lut-opt}
\end{figure*}

\subsection{LUT Acceleration with CMux Gate-level Parallelism}
\label{sec:lut}

\textbf{Conventional LUT evaluation suffers from low hardware utilization, elevated launch overhead, and redundant data access.}
As illustrated in Figure~\ref{fig:lut-opt}(a), the traditional approach employs slot-level parallelism to independently process each TFHE LWE ciphertext, requiring dedicated GPU resources for LUT evaluation (i.e., gate bootstrapping). 
However, gate bootstrapping typically involves multiple serial CMux operations based on \(n_{\rm{lwe}}\), with each operation relying on core operators such as \texttt{Decompose}, \texttt{NTT/INTT}, and \texttt{MAC}.
In this context, several issues arise:  
\ding{172} 
The core operators of the CMux gate accelerate only a single LWE ciphertext. 
However, small TFHE parameters restrict ciphertext size, leading to insufficient workload and low GPU utilization.
\ding{173} 
Gate bootstrapping involves \(n_{\rm{slot}} \cdot n_{\rm{lwe}}\) CMux operations executed serially on \(n_{\rm{slot}}\) TFHE LWE ciphertexts. 
Moreover, each CMux operation necessitates launching a separate kernel, dramatically increasing kernel launches.
\ding{174} 
Each CMux gate accesses the bootstrapping key for \texttt{external product} operations during the iterative process. 
Since all TFHE ciphertexts share the same key, it results in redundant accesses and memory inefficiencies.

\textbf{Propose CMux gate-level parallelization method to accelerate LUT evaluation.}
We observe that although \(n_{\rm{lwe}}\) TFHE ciphertexts contain different data, they share the same gate bootstrapping operations and bootstrapping keys across iterations.
Based on this, the CMux gate of all TFHE ciphertexts in each iteration can be merged into a single \textit{batched} CMux gate.
Specifically, as shown in Figure \ref{fig:lut-opt}(b), during the first iteration, all ciphertexts are fed into the batched CMux gate, using the bootstrap key RGSW($s_0$) as the selection bit. 
This method enables core operators (e.g., \texttt{Decompose}) to process all ciphertext data in a batch manner. 
As a result, the number of CMux executions and kernel launches is significantly reduced, minimizing repeated access to the bootstrapping key RGSW($s_i$).
In addition, the data processed by TFHE core operators increases by \(n_{\rm{lwe}}\) times, significantly enhancing hardware utilization. 
Table \ref{tab:lut-opt} shows the complexity of LUT evaluation with two levels of parallelism, indicating that our method reduces the number of executed CMux gates and accessed bootstrapping keys by a factor of \(n_{\rm{slot}}\). 
In contrast, the number of TFHE ciphertexts processed by each CMux gate increases by the same factor.

\begin{table}[t]
\setlength{\abovecaptionskip}{0.1cm}
\def\arraystretch{1.2}%
\centering
\scriptsize
\caption{Complexity analysis of LUT acceleration with two levels of parallelism.}
\begin{tabular}{|c|c|c|c|}
\hline
\textbf{Methods} & \textbf{\# CMux gate} & \textbf{\# Bootstrapping keys} & 
\textbf{\# TFHE ciphertext} \\ \hline
Slot & $n_{\rm slot} \cdot n_{\rm lwe}$ & $n_{\rm slot} \cdot n_{\rm lwe}$ & 1 \\ \hline
CMux gate & $n_{\rm lwe}$ & $n_{\rm lwe}$ & $n_{\rm slot}$ \\ \hline
\end{tabular}
\label{tab:lut-opt}
\vspace{-0.3cm}
\end{table}

\subsection{Repack Acceleration with Homomorphic Rotation-free MatVec Optimization}
\label{sec:repack}

\textbf{Existing matrix-vector multiplication necessitates numerous homomorphic rotations, greatly limiting repack performance.}
After LUT evaluation on the TFHE side, {repack}\cite{9519408,boura2020chimera,bae2023hermes,chen2021efficient,liu2023efficient} is performed to convert TFHE LWE ciphertexts into CKKS RLWE ciphertext in \textit{slot} format.
It consists of two steps: homomorphic evaluation of partial LWE decryption and homomorphic modulo \(q\) computation. 
The former involves a linear transformation (denoted as \texttt{LT}), specifically a matrix-vector multiplication (denoted as \textit{MatVec}) of the form \(\textbf{As} + \textbf{b}\).
Here, each row of the matrix \(\textbf{A}\), which has dimensions \(n_{\rm{slot}} \times n_{\rm{lwe}}\), represents an LWE ciphertext, while the vector \(\textbf{s}\) represents the RLWE encryption of the secret key \(\textit{sk}\), sized \(n_{\rm{lwe}} \times 1\).
Note that \textit{Tiling} technique is usually utilized to transform \(\textbf{A}\) into a square matrix (i.e., \(n_{\rm{slot}} = n_{{\rm lwe}}\)). 
There are two typical implementations for matrix-vector multiplication.
\ding{172} Typical {diagonal} method first transforms the plaintext matrix \(\textbf{A}\) into several diagonal vectors \({\bf u}_i\). 
Automorph is then applied to the ciphertext \(\textbf{s}\), followed by scalar multiplication and accumulation with these diagonal vectors (denoted as D-MatVec).
{\ding{173} Pegasus adopts the Baby-Step-Giant-Step (BSGS) algorithm to reduce the number of homomorphic rotations from \(n_{\rm{slot}} - 1\) to \(\sqrt{n_{\rm{slot}}}\), which in turn decreases the required rotation keys. However, as shown in Figure \ref{fig:peg-per}, \texttt{LT} in Pegasus still takes up to 35 seconds, accounting for 89\% of the total time. Therefore, optimizing linear transformation is crucial for improving repack performance.

\textbf{Propose a homomorphic rotation-free MatVec implementation to boost linear transformation.}
From the above analysis, the performance of \textit{MatVec} mainly depends on homomorphic rotations, as scalar multiplication incurs minimal overhead. 
Thus, the number of homomorphic rotations is a crucial factor limiting \texttt{LT} performance.
To tackle this issue, we propose a homomorphic rotation-free MatVec implementation that more effectively trades space for time. 
Specifically, in the BSGS-based MatVec implementation, Pegasus pre-stores all rotation ciphertexts during the \textit{Baby} step, enabling inexpensive scalar multiplications and requiring homomorphic rotations only in the \textit{Giant} step (P-MatVec). 
While this method aims to reduce the number of homomorphic rotations compared to the diagonal method, it still exhibits inefficiencies. 
Since the LWE secret key is known, it is feasible to precompute all rotation ciphertexts, thereby eliminating the need for real-time homomorphic rotations.
Furthermore, automorph inherently supports linear operations, including addition, multiplication, and the automorph itself. 
Leveraging this advantageous property, the automorph from the \textit{Giant} step can be combined with linear operations from the \textit{Baby} step.
It ultimately enables us to achieve a homomorphic rotation-free BSGS implementation (HRF-MatVec) that relies exclusively on inexpensive scalar multiplications. 
Figure \ref{fig:repack-opt} shows the workflow of the proposed {repack} implementation. 
When constructing the plaintext matrix of \(n_{\rm{slot}}\) TFHE LWE ciphertexts, we classify the matrix as \textit{fat} or \textit{thin} based on its dimensions, which determines whether to apply {row tiling} or {column tiling} for dimensional expansion. 
Next, the diagonal vectors are grouped with precomputed \(n_{\rm{slot}}\) rotation ciphertexts according to the \textit{Giant} step and \textit{Baby} step. 
Finally, scalar multiplications are performed, and the results are accumulated to generate the CKKS RLWE ciphertext. 
Table \ref{tab:repack} details the complexity of {repack} under different methods. 

\begin{table}[t]
\setlength{\abovecaptionskip}{0.1cm}
\def\arraystretch{1.2}%
\centering
\begin{center}
\scriptsize
\caption{{Complexity analysis of {repack} acceleration with different matrix-vector multiplication implementations.}}
\setlength{\tabcolsep}{1.7mm}{
\begin{tabular}{|m{1.7cm}<{\centering}|m{1.2cm}<{\centering}|m{1.5cm}<{\centering}|m{1.2cm}<{\centering}|m{1.3cm}<{\centering}|}  
\hline  
\textbf{Implementations} & \textbf{\# Rotation \newline Operations} & \textbf{\# Scalar \newline Multiplication} &\textbf{\# Rotation \newline Ciphertext} & \textbf{\# Rotation \newline Keys} \\ \hline  
D-MatVec & $ {n_{\rm slot}}$ & ${n_{\rm slot}}$ & $1$ & ${n_{\rm slot}}$\\ \hline  
% P-BSGS &$n_1+n_2$ & $n_1\times n_2$ & $1$ & $n_1+n_2$\\ \hline    
P-MatVec & $\approx \sqrt{n_{\rm slot}}$ & ${n_{\rm slot}}$ &$\approx \sqrt{n_{\rm slot}}$ & $\approx \sqrt{n_{\rm slot}}$\\ \hline  
HRF-MatVec & $0$ & ${n_{\rm slot}}$ & ${n_{\rm slot}}$ & $0$\\ \hline
\end{tabular}}
\label{tab:repack}
\end{center}
\vspace{-0.8cm}
\end{table}

\begin{figure*}[t]
\setlength{\belowcaptionskip}{-0.6cm}
\setlength{\abovecaptionskip}{0cm}
\centering
\includegraphics[width=\linewidth]{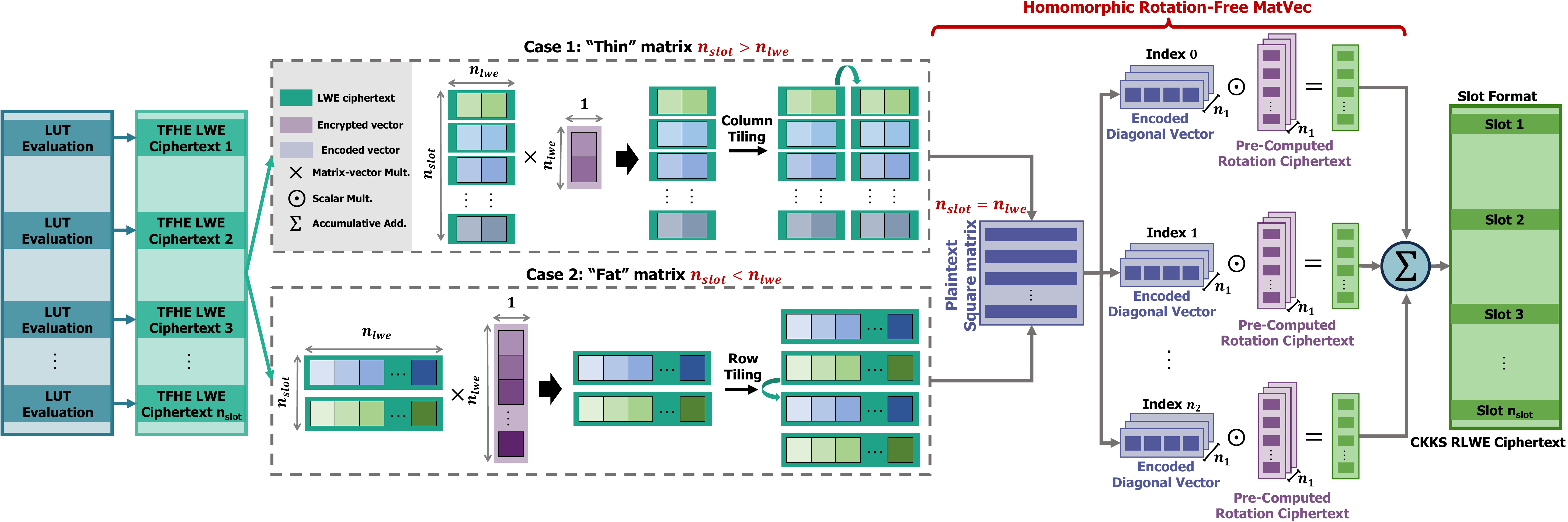}
\caption{Repack acceleration with homomorphic rotation-free MatVec optimization.}
\label{fig:repack-opt}
\end{figure*}

\section{{Scalable NTT Acceleration Design on GPUs}}
\vspace{-0.15cm}

In this section, we propose tailored optimization techniques for polynomials of varying lengths in CKKS and TFHE, specifically targeting synchronization bottlenecks.
First, the staged structure of NTT requires thread synchronization at the end of each stage, leading to excessive synchronization. 
Second, synchronization frequently occurs across thread blocks, necessitating global synchronization and imposing substantial delays. 
Third, although hybrid synchronization techniques can alleviate some overhead, they do not fully capitalize on the optimal switching point.
To address these challenges, we contribute a series of strategies:
\textit{(\rmnum{1}) To reduce the number of synchronization}, we propose two NTT optimization designs based on stage fusion: butterfly decomposition method for small TFHE polynomials and thread aggregation method for large CKKS polynomials.
\textit{(\rmnum{2}) To minimize synchronization scale}, we introduce a polynomial coefficient shuffling technique that significantly reduces the costs associated with global synchronization.
\textit{(\rmnum{3}) To identify the optimal synchronization switching point}, we propose an SM-aware synchronization strategy, which balances synchronization overhead with hardware utilization.
The following sections provide an in-depth analysis of these optimizations.

\vspace{-0.15cm}
\subsection{Butterfly Decomposition-based NTT for TFHE Polynomial}
\label{sec:butterfly}
\vspace{-0.15cm}

In traditional GPU-based radix-2 NTT, each thread processes a {complete} butterfly as the fundamental unit. 
This approach causes one computation result to serve as input for another thread in the next stage, leading to excessive thread stalls and performance degradation. 
We observe that \textit{a complete butterfly can be decomposed into finer-grained units, termed half-butterflies}. 
By reallocating these units, stage fusion becomes possible, effectively eliminating the need for explicit thread synchronization. Building on this observation, we propose the butterfly decomposition method.

\textbf{Decompose complete butterflies to fuse successive stages.}
As shown in Figure \ref{fig:ntt-decomp}, in a typical 4-point NTT, threads 0 and 1 perform a \textit{complete} butterfly operation at two stages, $S_A$ and $S_B$, respectively. 
Synchronization is required between the two threads as the output of $S_A$ from one thread is used as input for $S_B$ in the other thread, and vice versa.
Using butterfly decomposition, each thread processes two half-butterflies (e.g., the upper triangles for thread 0 and the lower triangles for thread 1) from two different complete butterflies at stage $S_A$. 
The outputs then serve as inputs for the same thread at stage $S_B$, eliminating the need for synchronization between threads. 
Stages $S_A$ and $S_B$ are fused into a single stage, $S_{AB}$. 
This approach further reduces synchronizations by decomposing multiple butterflies while maintaining the same number of threads ($N/2$) as radix-2 NTT, ensuring parallelism and suitability for small TFHE polynomials.
Figure \ref{fig:sf-radix2} compares the execution times of traditional radix-2 NTT and butterfly decomposition-based NTT.  The proposed method reduces the number of synchronizations for a 16-point NTT from 3 to 1, resulting in remarkable performance gains.

\begin{figure}[t]
\setlength{\belowcaptionskip}{-0.7cm}
\setlength{\abovecaptionskip}{0cm}
\centering
\includegraphics[width=0.9\linewidth]{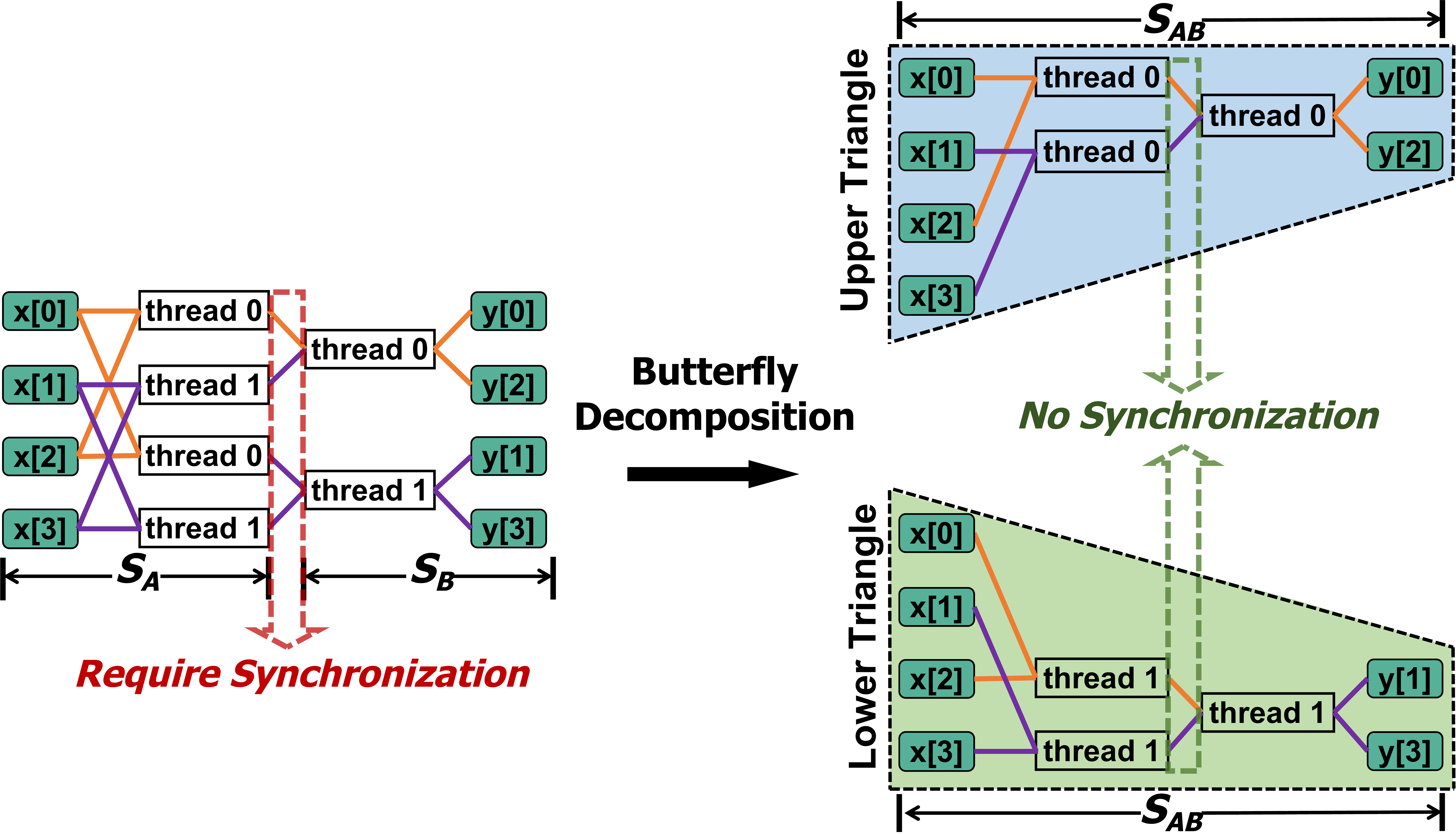}
\caption{Butterfly decomposition principle. The calculation of the decomposed same-color area is completed by the same thread without synchronization.} 
\label{fig:ntt-decomp}
% \vspace{-0.2cm}
\end{figure}

\textbf{Employ polynomial replicas to address data conflicts.}
Unfortunately, the butterfly decomposition method also brings about potential data conflict issues. 
As shown in Figure \ref{fig:ntt-decomp}, threads 0 and 1 may simultaneously read from or write to the same data points, resulting in \textit{spatial} conflicts. 
Moreover, when thread 1 reads the data after thread 0 writes the data pair x[0]/x[1] in an in-place manner, \textit{read-after-write} (RAW) conflicts occur. 
These conflicts can cause significant data inconsistencies and ultimately lead to failures in NTT conversion.
To resolve these issues, we introduce a polynomial replica that tracks intermediate results and ensures consistency with the original polynomial during computation. 
Using this technique, memory mapping remains efficient and conflict-free. When entering a new stage, threads computing the upper or lower triangle of the same butterfly access the polynomial vector {{x} and its replica {x\_copy}}, respectively. 
For polynomials of any length with an odd number of stages, the remaining stages can utilize the radix-2 implementation.

\vspace{-0.15cm}
\subsection{Thread Aggregation-based NTT for CKKS Polynomial}
\label{sec:ntt_highradix}
\vspace{-0.15cm}

When dealing with larger CKKS polynomials, the butterfly decomposition method may encounter performance degradation due to potential thread divergence. 
Hence, it is imperative to investigate further an optimization method suitable for such cases while reducing the number of synchronizations.
We observe that \textit{it is possible to aggregate two threads responsible for different butterfly units into a single thread.}
This aggregation enables stage fusion, allowing a single thread to handle multiple butterflies simultaneously. 
With this strategy, we propose the thread aggregation method.

\begin{figure}[t]
\setlength{\belowcaptionskip}{-0.2cm}
\setlength{\abovecaptionskip}{0.1cm}
\centering
\includegraphics[width=0.9\linewidth]{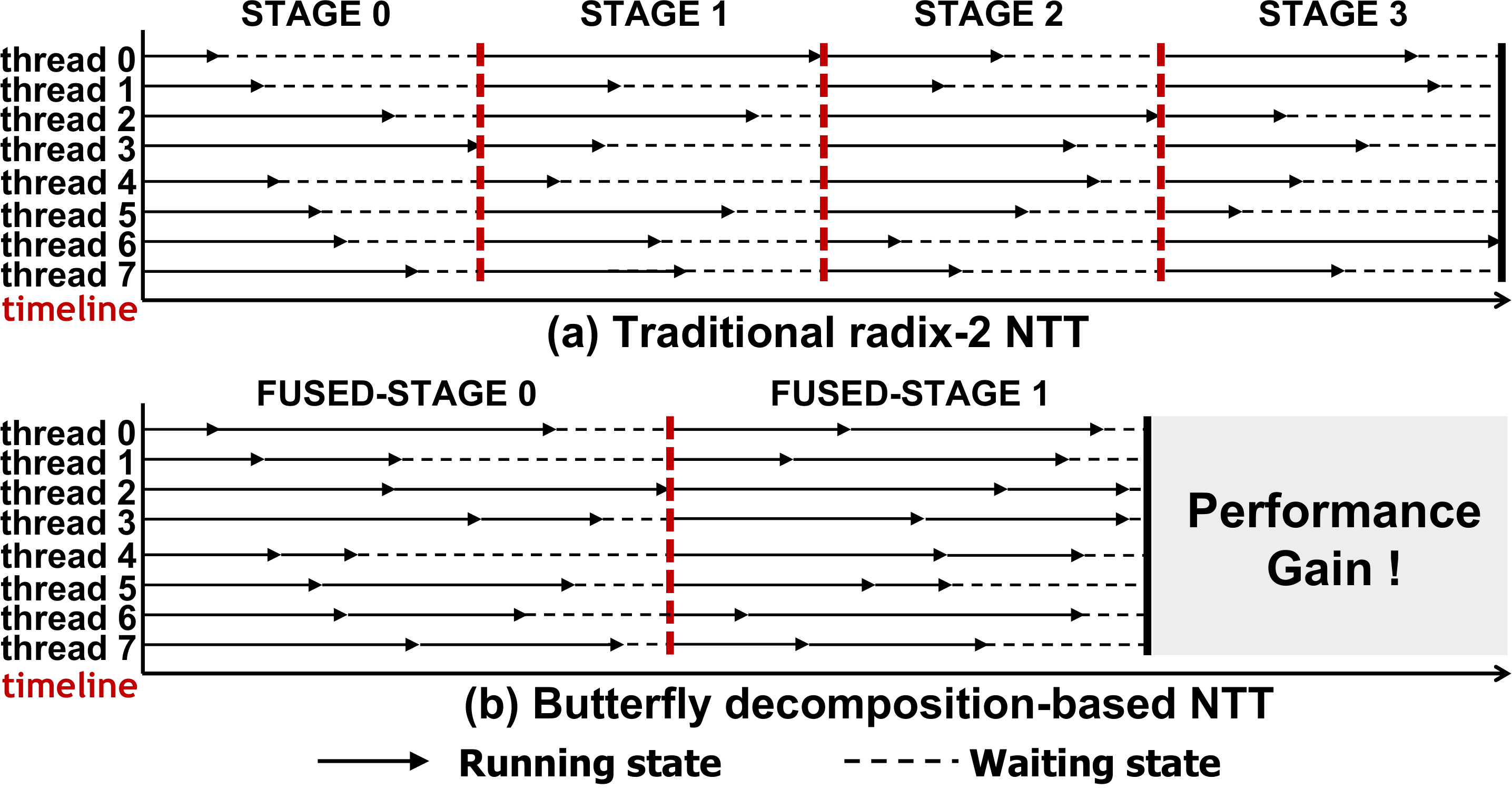}
\caption{16-point NTT execution time comparison between radix-2 NTT and butterfly decomposition-based NTT.} 
\label{fig:sf-radix2}
% \vspace{-0.2cm}
\end{figure}

\textbf{Aggregate multiple threads to achieve stage fusion.}
Figure \ref{fig:ntt-aggre} illustrates the thread aggregation principle. 
In a four-point NTT, two threads process two butterfly units per stage, with synchronization needed for data consistency in stage 1. 
Using the thread aggregation method, the butterfly units {x[1]/x[3]} and {x'[2]/x'[3]}, initially handled by thread 1, are assigned to thread 0.
Thread 0 executes all butterflies in stage 0 and generates all inputs for stage 1, allowing the two stages to be fused into one.
Significantly, this design can scale by allowing one thread to compute multiple butterfly units, further aggregating threads and reducing the number of stages. Figure \ref{fig:radix2-aggre} shows the execution time comparison between radix-2 NTT and thread aggregation-based NTT, highlighting a significant performance boost by reducing the number of threads from 8 to 4 in principle.

\textbf{Utilize loop reconstruction and redirection index to address data mapping challenges.}
In this context, two challenges arise:
(\rmnum{1}) The loop structure of traditional radix-2 NTT limits thread aggregation by requiring each thread to process only one butterfly unit.
(\rmnum{2}) Careful mapping of these units onto the GPU is essential, especially when multiple butterfly units are assigned to a single thread.
To address these issues, we follow the principle that the number of butterfly units contained in an $N$-point NTT remains invariant, i.e., $\frac{N}{2}\cdot{\rm log}_2N$.
It can be deduced that the relationship between the number of threads $K$ and the number of butterfly units $H$ processed by each thread satisfies $K \cdot H \cdot {\rm log}_{2^A}N = \frac{N}{2}\cdot{\rm log}_2N$, 
where $A$ denotes the number of aggregated threads, and $H$ equals $2^A$. 
For example, if $A = 2$, two threads are aggregated to form $\log_{2^2} N$ stages. 
In this case, each thread processes four butterfly units and the number of threads $K$ is set to $N/4$.
We further introduce a fresh \textit{redirection index} to map each thread to multiple butterfly units at each stage and to identify the relevant twiddle factors. Like the butterfly decomposition, this method also supports cases where the number of stages is not divisible by 2.

\begin{figure}[t]
\setlength{\belowcaptionskip}{-0.5cm}
\setlength{\abovecaptionskip}{0.1cm}
\centering
\includegraphics[width=0.9\linewidth]{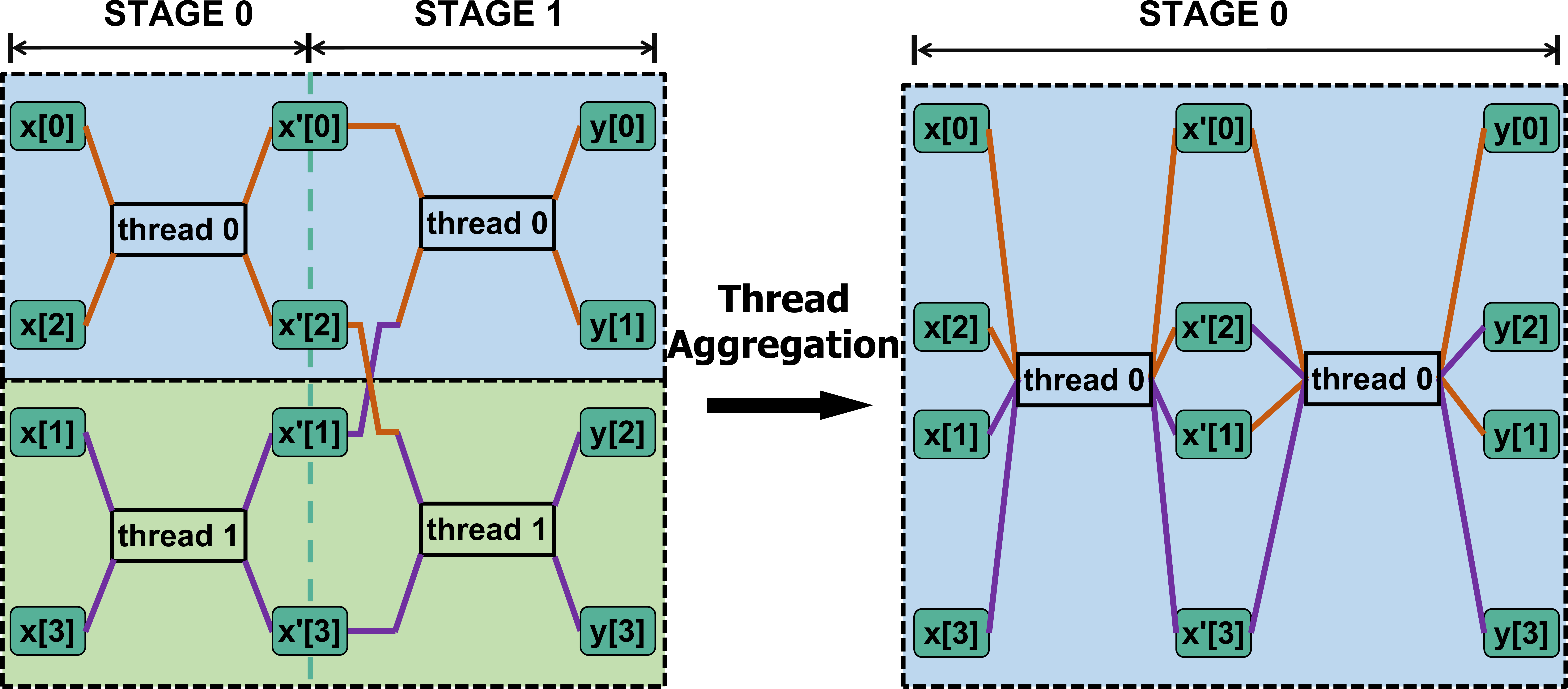}
\caption{{Thread aggregation principle.} The calculation of the same-color area is completed by the same thread, and lines of the same color make up a butterfly.}
\label{fig:ntt-aggre}
\end{figure}

\begin{figure}[t]
\setlength{\belowcaptionskip}{-0.6cm}
\setlength{\abovecaptionskip}{0.1cm}
\centering
\includegraphics[width=0.9\linewidth]{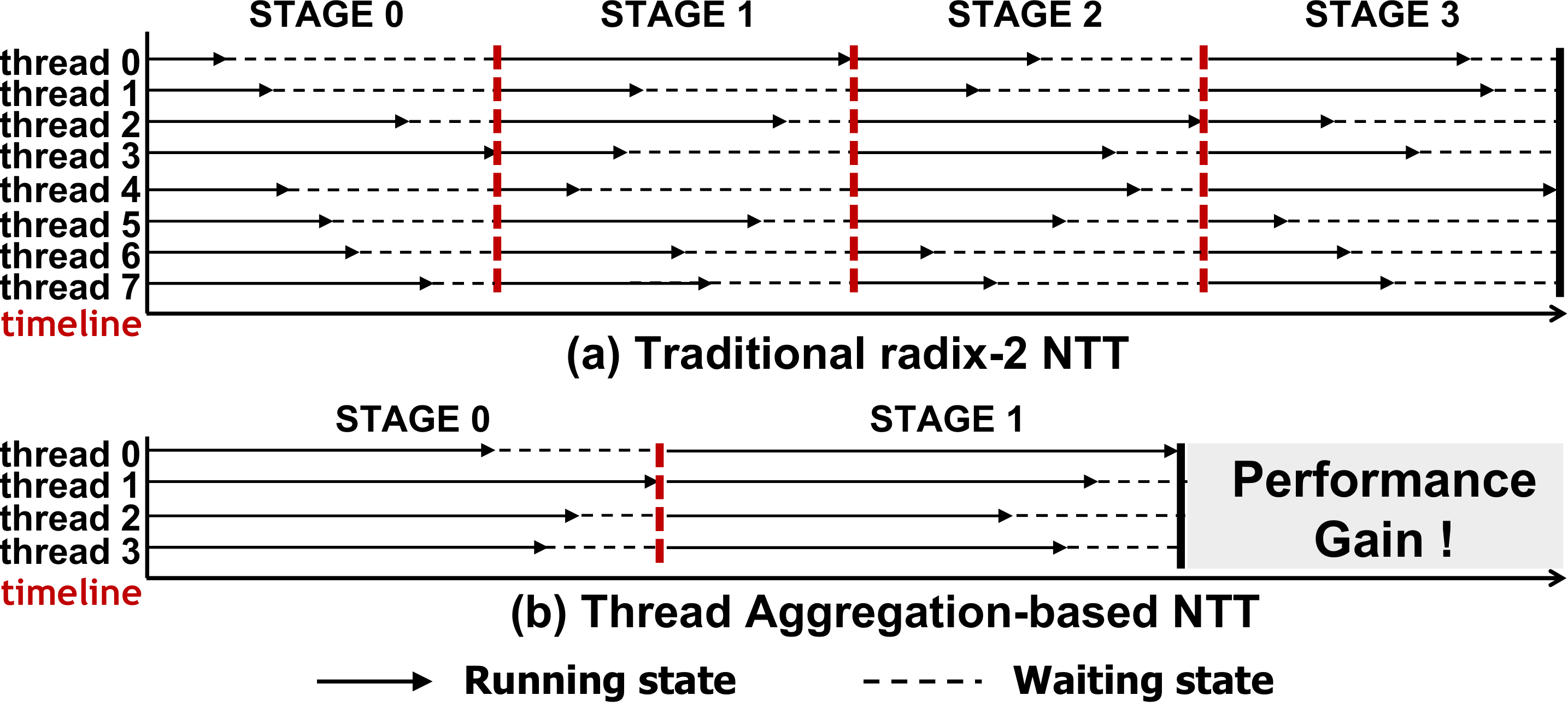}
\caption{16-point NTT execution time comparison between radix-2 NTT and thread aggregation-based NTT.} 
\label{fig:radix2-aggre}
\end{figure}

\begin{figure*}[t]
\setlength{\belowcaptionskip}{-0.6cm}
\setlength{\abovecaptionskip}{0.1cm}
\centering
\includegraphics[width=\linewidth]{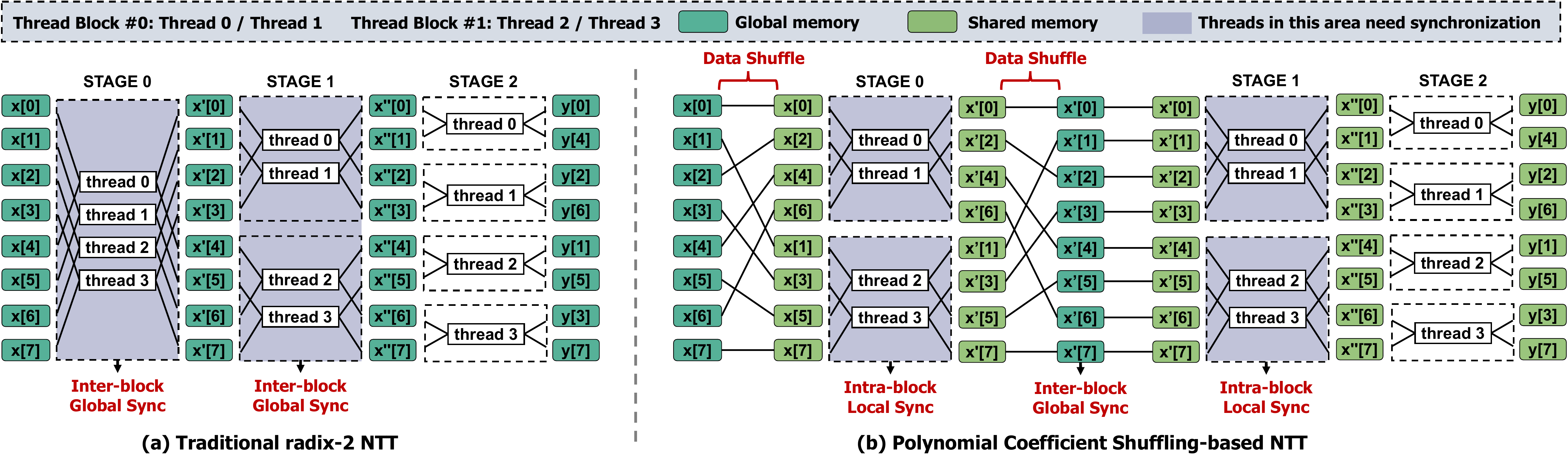}
\caption{Traditional radix-2 NTT and polynomial coefficient shuffling-based NTT.}
\label{fig:ntt-intra}
\end{figure*}

\subsection{Polynomial Coefficient Shuffling-based NTT}
\label{sec:pcs}

Despite the proposed methods to reduce the number of synchronizations, current NTT implementations still face significant challenges, including expensive global synchronization overhead and inefficient memory utilization.
First, conventional methods distribute polynomial coefficients across multiple thread blocks. Still, when dependent coefficients are in different blocks, global synchronization is required, forcing threads to wait for each other and causing long latency. 
Second, storing polynomials in global memory for shared access reduces memory efficiency due to the long transfer path between global memory and thread registers.

A simple solution to these issues is to load the entire polynomial into a single thread block's shared memory\cite{ozerkefficient}. However, this severely limits GPU hardware resource utilization, and the shared memory capacity is insufficient to store 64-bit polynomials of length 8192 or larger. 
Thus, \textit{it is essential to explore strategies for distributing polynomial coefficients across the shared memory of multiple thread blocks while eliminating the need for inter-block global synchronization}.

\textbf{Convert inter-block global synchronization into local synchronization within multiple independent thread blocks.}
We observe that when interdependent data is distributed across multiple thread blocks, reorganizing the data layout to consolidate it within a single thread block can offer two key benefits: First, it reduces global synchronization overhead by limiting it to the intra-block level.
Second, it enhances memory access efficiency by leveraging the faster speeds of shared memory, minimizing long-distance data transfers. Thus, devising an efficient data allocation strategy that restricts synchronization within individual thread blocks is crucial for optimizing performance.

\textbf{Shuffle polynomial coefficient to eliminate data dependencies between thread blocks.}
As shown in Figure~\ref{fig:ntt-intra}(a), in traditional methods, when four GPU threads perform an 8-point NTT, each thread handles two elements per stage. In Stage 0, thread 0 processes {x[0]/x[4]}, and thread 2 processes {x[2]/x[6]}. In Stage 1, thread 0 then processes {x[0]/x[2]} (from thread 2), and thread 2 processes {x[6]/x[4]} (from thread 0). 
This process induces a dependency between thread 0 and thread 2 across different blocks, requiring global synchronization.
As shown in Figure~\ref{fig:ntt-intra}(b), shuffling the polynomial coefficients before Stage 0 rearranges the data in global memory, ensuring that dependent data resides in the same thread block’s shared memory. 
For instance, in thread block 0, thread 0 and thread 1 process {x[0]/x[2]/x[4]/x[6]}. After completing the butterfly operations, only local synchronization within the thread block is required.
It is important to note that while one global synchronization is still necessary before entering Stage 1 for data consistency, this approach dramatically reduces the number of global synchronizations from a maximum of 11 (in the radix-2 setting) to just 1 for $N$ equals $4096$, compared to traditional methods. Moreover, the polynomial coefficient shuffling technique works independently of the stage fusion optimization, and their integration can further enhance performance.

\begin{figure}[t]
\setlength{\belowcaptionskip}{-0.6cm}
\setlength{\abovecaptionskip}{0.1cm}
\centering
\includegraphics[width=0.9\linewidth]{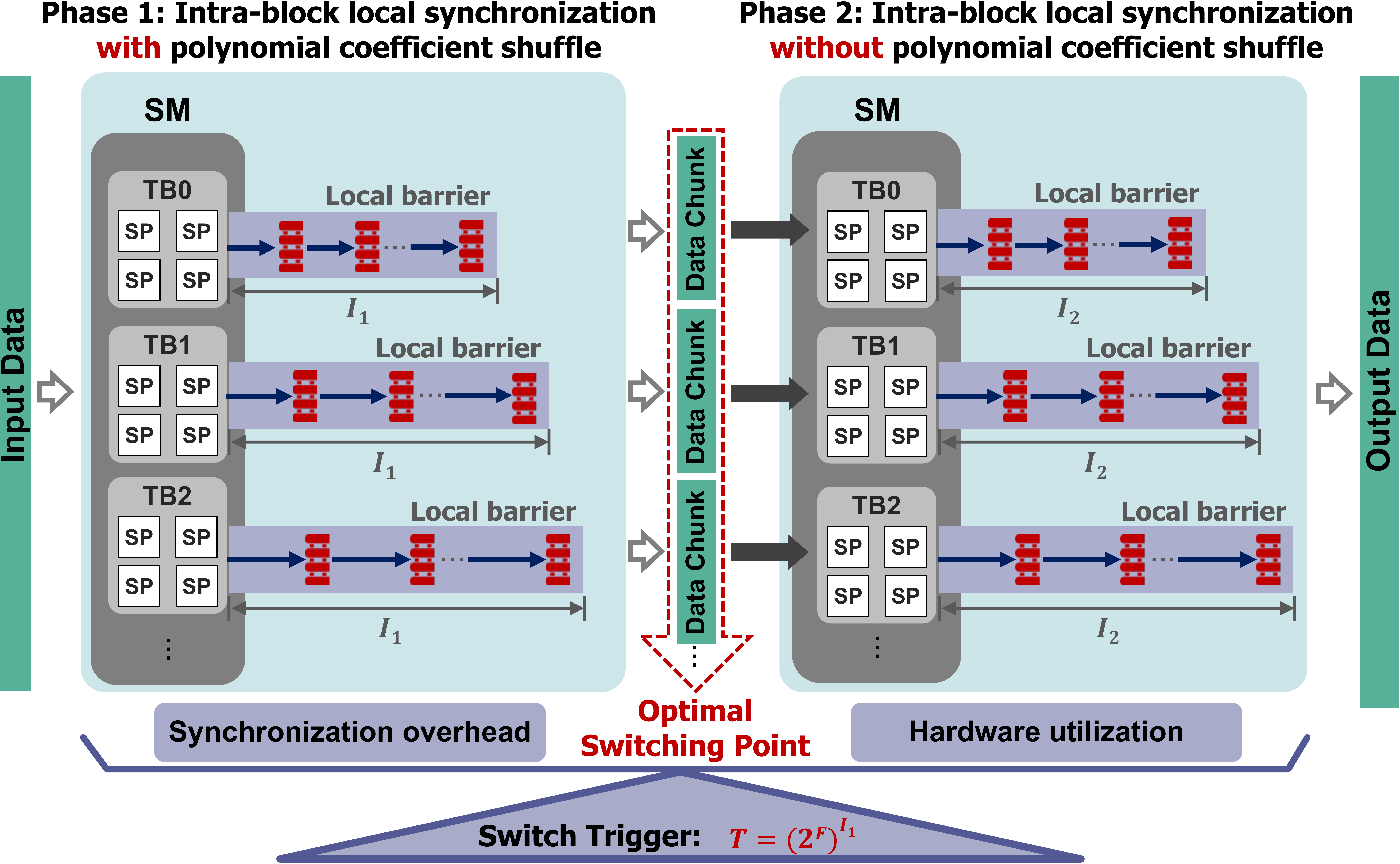}
\caption{Optimal synchronization switching point-based NTT using the switch trigger $T=(2^F)^{I_1}$ that balances synchronization overhead and hardware utilization.}
\label{fig:ntt-hh}
% \vspace{-0.5cm}
\end{figure}

\vspace{-0.15cm}
\subsection{{Optimal Synchronization Switching Point-based NTT}}
\label{sec:ntt_synclevel}
\vspace{-0.15cm}

A typical hybrid implementation involves inter-block global synchronization (Phase 1) and intra-block local synchronization (Phase 2). Applying the techniques from Section~\ref{sec:pcs}, the global synchronization in Phase 1 can be replaced with local synchronization based on polynomial coefficient shuffling. It is important to note that Phase 2 handles independent data blocks, which do not require this approach. Additionally, although both phases employ the same synchronization strategy, they cannot be merged, as each thread block processes data from consecutive stages rather than all stages simultaneously.
Furthermore, the position of the switching point between these two phases is crucial for balancing synchronization overhead and hardware utilization. A key question arises: "\textit{How can we identify the optimal synchronization switching point?}"

\textbf{Previous efforts switched too early to minimize synchronization overhead, but hardware utilization drops sharply, leading to suboptimal performance.}
Specifically, the synchronization point is set to switch when an independent data chunk of 2048 elements is reached (based on the maximum of 1024 threads). However, this method overlooks the fact that only a limited number of thread blocks can be scheduled to SMs, resulting in low hardware utilization. On the other hand, switching {too late} would cause excessive synchronization overhead, even if enough data blocks were allocated to SMs to ensure full hardware utilization.

\textbf{Explore SM-aware synchronization combination strategy to balance synchronization overhead and hardware utilization.}
During NTT execution, the data sequence is split into independent data chunks (denoted as $D$), which require an equal number of thread blocks (denoted as $T$), satisfying the condition $T=D$. 
The number of data chunks after each stage can be predicted, allowing the required thread blocks to be determined in advance. 
However, a challenge arises when the number of streaming multiprocessors (SMs) on the GPU does not match $T$ (typically a power of 2). 
For example, the NVIDIA Tesla A100 GPU supports 108 SMs. 
In such cases, it's crucial to optimize $T$ for maximum hardware utilization and minimal synchronization overhead.

Based on these observations, we can use the following switch trigger to find the optimal switching point: 
{Given the number of SMs, $S$, we explore all synchronization combinations $(I_1, I_2)$ and determine the optimal point by checking if $T=(2^F)^{I_1}$ is closest to $S$ for a given $I_1$.} The total number of synchronizations is $I = I_1 + I_2 = \log_{2^F} N - 1$, where $I_1$ refers to intra-block synchronizations \textit{with} polynomial coefficient shuffling, $I_2$ refers to those \textit{without} polynomial coefficient shuffling, and $F$ represents the number of fused stages based on butterfly decomposition and thread aggregation.
It's important to note that $T$ can vary across GPUs with different computing capabilities due to differences in the number of SMs. This aspect is further examined in the experiment section.
Figure \ref{fig:ntt-hh} illustrates the {optimal synchronization switching point-based} NTT.
In Phase 1, intra-block local synchronizations using the polynomial coefficient shuffling method split the data sequence until $I_1$ reaches the trigger point. In Phase 2, the resulting data chunks are loaded into thread blocks, each carrying out $I_2$ local synchronizations.

\textbf{Develop an automated kernel configuration technique to dynamically determine the optimal switching point.}
Once the polynomial length and the number of SMs have been chosen, the NTT kernel configuration can be automatically generated. 
For instance, after determining the optimal switching point $T=(2^F)^{I_1}$, the number of thread blocks can be calculated accordingly, equivalent to the number of data chunks. 
Similarly, the number of elements in each data chunk, represented as $H$, can be obtained by computing $H=N/T$.
In addition, the number of threads assigned to each thread block represented as $E$ can be determined using $E_{\tt BD}=\frac{N}{({2}^{F \times I_1}) \times 2}$ and $E_{\tt TA}=\frac{N}{({2}^{F \times I_1}) \times R}$, where the {butterfly decomposition} method maintains a consistent number of threads. 
In contrast, the {thread aggregation} varies depending on the number of threads aggregated, denoted as $R$.

\vspace{-0.15cm}
\section{Evaluation}
\label{sec:evaluation}
\vspace{-0.15cm}
\subsection{Methodology} 
\textbf{Software and hardware configurations.} 
{The experiments are conducted on a realistic GPU server equipped with an Intel i9-10900K CPU (10 cores, 3.7 GHz, 128 GB DRAM) and an NVIDIA GeForce RTX 3070 GPU (46 SMs, 5888 cores, 1.5 GHz). 
A NVIDIA Tesla A100 GPU (108 SMs, 6912 cores, 1.41 GHz) is also used as an alternative platform. 
The operating system is Ubuntu 18.04.5, and our implementation is built using CUDA Toolkit 11.2. 
Compiler tools include GCC 9.0.0, CMake 3.19.3, and NVCC 11.2.67. NVIDIA Nsight Compute Tools are used for kernel profiling. }

\begin{figure*}[h!]
\setlength{\abovecaptionskip}{0.1cm}
\centering
\includegraphics[width=\linewidth]{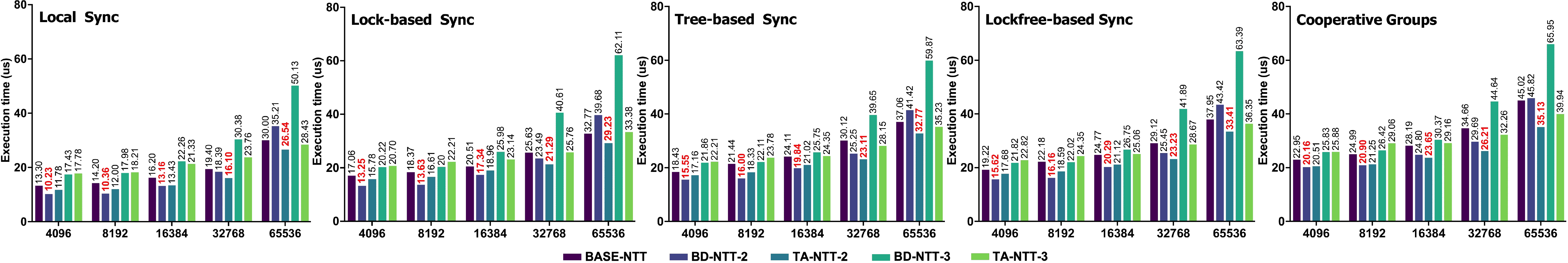}
\caption{Performance of NTT implementation based on {butterfly decomposition} and thread aggregation methods.} 
\label{fig:ntt}
\end{figure*}

\begin{figure*}[h!]
\setlength{\belowcaptionskip}{-0.3cm}
\centering
\includegraphics[width=\linewidth]{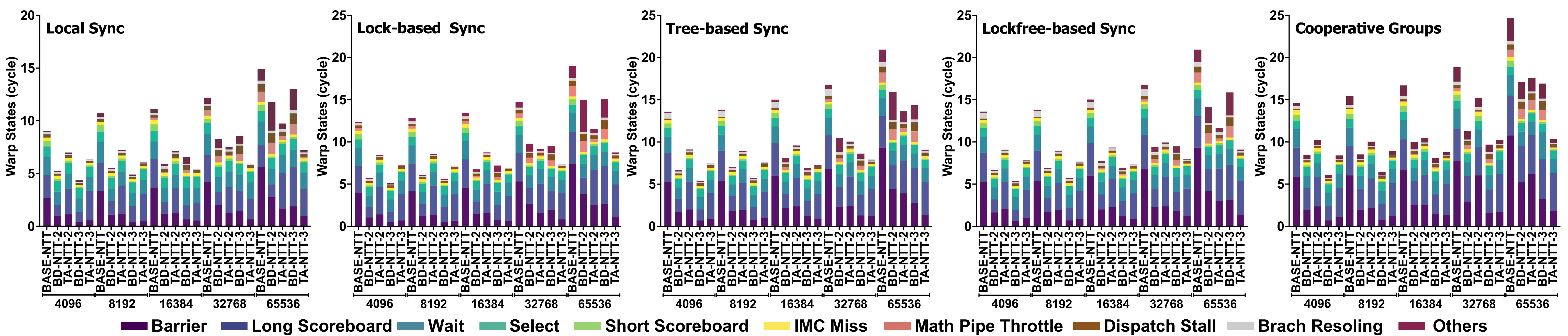}
\caption{Breakdown of warp states through profiling NTT kernel under different optimizations.} 
\label{fig:ntt-stall}
\end{figure*}

\textbf{Benchmarks.} 
Specifically, we set the polynomial length range from 4096 to 65536, and $q$ is set to a typical double-word size of 56 bits.
Five typical kernel-inside synchronization methods are employed~\cite{xiao2010inter,wang2023he}.
Note that the proposed solutions are entirely orthogonal to any emerging synchronization techniques on future high-end GPUs.
The performance of the {butterfly decomposition-based} NTT ({\tt BD-NTT}), {thread aggregation-based} NTT ({\tt TA-NTT}), polynomial coefficient shuffling-based NTT ({\tt PCS-NTT}) and {optimal synchronization switching point-based} NTT ({\tt OSSP-NTT}) are evaluated, respectively.
In particular, {\tt BD-NTT-\#} and {\tt TA-NTT-\#} indicate the number of fused stages.
Moreover, we take the radix-2 NTT as the baseline and denote it as {\tt BASE-NTT}.

For CKKS primitives and bootstrapping, we benchmark our implementation against three leading open-source CPU libraries: {HElib}~\cite{halevi2014algorithms}, {SEAL}~\cite{sealcrypto}, and {OpenFHE}~\cite{al2022openfhe}, as well as cutting-edge GPU acceleration techniques: {Over100x}\cite{jung2021over}, {HE-Booster}\cite{wang2023he}, {Phantom}\cite{yang2024phantom} and {TensorFHE}\cite{fan2023tensorfhe}. 
For TFHE operations (e.g., gate bootstrapping), we compare our results with the {TFHE-rs} open-source library~\cite{TFHE-rs} and available GPU implementations (e.g., {cuFHE}\cite{cuFHE} and {nuFHE}\cite{NuFHE}).
Additionally, we conduct experiments to evaluate scheme switching performance and provide a detailed comparison against the state-of-the-art implementation, {Pegasus}~\cite{lu2021pegasus}.

\subsection{Performance Analysis of {Butterfly Decomposition} and {Thread Aggregation} Optimizations}

Figure \ref{fig:ntt} illustrates the execution time of a single NTT kernel.
The following observations can be concluded:

\begin{itemize}[leftmargin=0.5cm, noitemsep, nolistsep]
    \item 
    \textbf{Performance benefits of butterfly decomposition method for TFHE polynomial.}
    When $N$ is 8192, {\tt BD-NTT-2} outperforms {\tt BASE-NTT} by 1.37$\times$, 1.35$\times$, 1.34$\times$, 1.37$\times$, and 1.2$\times$ across five methods. 
    However, two key points are worth noting. 
    First, for $N = 65536$, {\tt BD-NTT-2} has longer delays than the baseline. Second, decomposing more butterflies does not continually improve performance. 
    For example, {\tt BD-NTT-3}, which decomposes two butterflies, performs the worst. 
    This suggests that large polynomials or excessive decomposition may cause thread divergence and degrade performance.
    \item 
    \textbf{Performance benefits of thread aggregation method for CKKS polynomial.}
    When \textit{N} is 32768, {\tt TA-NTT-2} achieves 1.2$\times$, 1.2$\times$, 1.3$\times$, 1.25$\times$, and 1.32$\times$ speedup over the baseline across the five synchronization strategies. 
    However, aggregating more threads does not always improve performance. 
    While it reduces global synchronizations, it also lowers the number of threads launched. 
    For example, {\tt TA-NTT-3}, with fewer synchronization needs, shows no performance gain due to its decreased threads, leading to decreased parallelism and utilization.
    \item 
    \textbf{Choose of synchronization method for different parameters.}
    \textit{The butterfly decomposition method outperforms the thread aggregation implementation for small TFHE polynomials. }
    For $N$ is 8192, {\tt BD-NTT-2} achieves speedups of 1.14$\times$, 1.22$\times$, 1.15$\times$, 1.15$\times$, and 1.02$\times$ over {\tt TA-NTT-2} across the five synchronization methods. 
    Conversely, \textit{thread aggregation offers greater speedup than butterfly decomposition for large CKKS polynomials.} 
    For instance, when $N$ is 65536, {\tt TA-NTT-2} achieves speedups of 1.33$\times$, 1.27$\times$, 1.26$\times$, 1.3$\times$, and 1.3$\times$ over {\tt BD-NTT-2}.
\end{itemize}

\textbf{Deep analysis of stalling cycles for different synchronization methods:}
Using kernel profiling, we further measure the warp states (i.e., average stalling cycles for instructions). 
Figure~\ref{fig:ntt-stall} shows the breakdown of warp states.

\begin{itemize}[leftmargin=0.5cm, noitemsep, nolistsep]
    \item 
    \textbf{Baseline NTT.} 
    The baseline (i.e., {\tt BASE-NTT}) has the most stalling cycles for instructions compared to other implementations. 
    This is due to the fact that it requires more inter-block global synchronizations.
    Therefore, correspondingly more stalling cycles are consumed.
    
    \item \textbf{{Butterfly decomposition-based} NTT for TFHE polynomial.}
    Although this method has fewer stalling cycles than other implementations, it is sometimes slower. 
    This is due to a significant increase in thread divergence: while the baseline averages 11.13 divergent branches for $N = 65536$, {\tt BD-NTT-2} and {\tt BD-NTT-3} have 27.83 and 51.48, respectively. 
    The drawback of thread divergence offsets the stalling cycle improvements, making the butterfly decomposition method less effective with larger polynomials.
    
    \item \textbf{{Thread aggregation-based} NTT for CKKS polynomial.}
    Aggregating more threads fuses more stages, reducing global synchronizations and stalling cycles. 
    However, despite having the fewest stalling cycles, {\tt TA-NTT-3} does not outperform {\tt BASE-NTT} and {\tt TA-NTT-2}. 
    This is because {\tt TA-NTT-3} using only \textit{N}/8 threads, which is just 1/4 and 1/2 of the threads in {\tt BASE-NTT} and {\tt TA-NTT-2}, respectively, leading to reduced parallelism.
    
\end{itemize}

\begin{figure}[t]
\setlength{\belowcaptionskip}{-0.7cm}
\setlength{\abovecaptionskip}{0.1cm}
\centering
\includegraphics[width=0.9\linewidth]{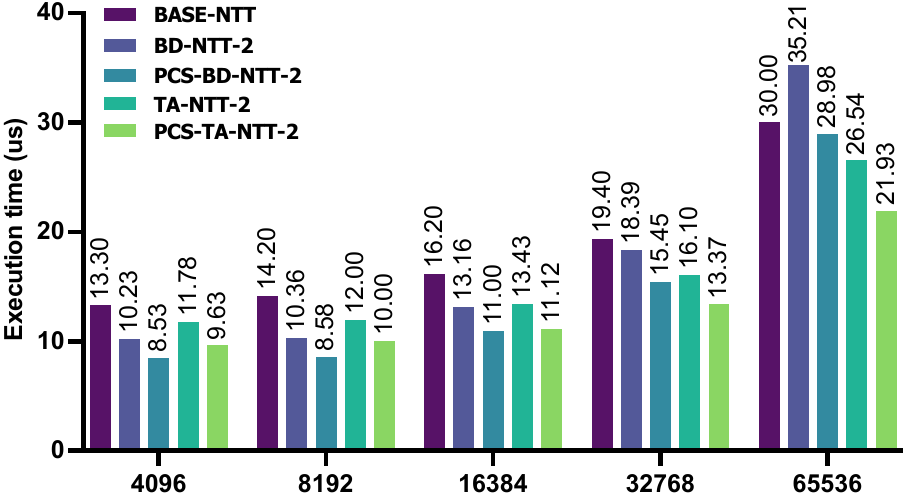}
\caption{Performance of polynomial coefficient shuffling-based NTT implementation with \textit{Local Sync} method.}
\label{fig:ntt-pcs}
% \vspace{-0.5cm}
\end{figure}

\subsection{Performance Analysis of Polynomial Coefficient Shuffling-based NTT}

As shown in Figure \ref{fig:ntt-pcs}, we evaluate the performance of the polynomial coefficient shuffling method. 
We also integrated butterfly decomposition and thread aggregation as stage fusion methods to reduce the number of synchronizations.
The results show that the polynomial coefficient shuffling method significantly boosts performance by optimizing GPU shared memory for faster data access and reducing global synchronization overhead. 
This is accomplished by replacing time-consuming global synchronization with smaller intra-thread block synchronization.
As a result, when \(N=4096\), the performance improved by 1.25 times ({\tt PCS-BD-NTT-2} vs. {\tt BD-NTT-2}), and at \(N=65536\), the performance improved by 1.2 times ({\tt PCS-TA-NTT-2} vs. {\tt TA-NTT-2}).
It further validates the effectiveness of the polynomial coefficient shuffling method when integrated with two stage fusion techniques.

\begin{table}[t]
\begin{center}
\def\arraystretch{1.1}%
\setlength{\abovecaptionskip}{0.1cm} 
\caption{Performance evaluation (microseconds) of the optimal synchronization switching point-based NTT.}
\label{tab:ntt-mix}
\begin{threeparttable}
\scriptsize
\centering
\setlength{\tabcolsep}{1.6mm}{
\begin{tabular}{cccccc}
\toprule
\multirow{2}{*}{\textbf{Method}}&
 \multicolumn{5}{c}{\textbf{Polynomial length \textit{N}}} \\
 \cmidrule(r){2-6} 
  & \textbf{4096} & \textbf{8192}  & \textbf{16384} & \textbf{32768} &  \textbf{65536} \\
\midrule
M1   & 10.23/9.59   & 10.36/9.87   & 13.16/12.33 & 18.39/16.99 & 35.21/32.36 \\
M2   & 8.21/7.78  & 8.83/7.95  & 10.96/10.16 & 15.72/14.23 & 29.78/27.88 \\
\textbf{M3}   & \textbf{7.14/6.82}  & \textbf{7.69/6.93}  & \textbf{9.61/8.87} & \textbf{13.54/12.47} & \textbf{25.89/24.31} \\
M4   & 11.78/11.02   & 12/11.1   & 13.43/12.78 & 16.1/15.03 & 26.54/24.17 \\
M5   & 9.74/9.21 & 10.23/9.56  & 11.76/10.89 & 13.75/13.02 & 22.87/21.3 \\
\textbf{M6}   & \textbf{8.2/7.19}  & \textbf{8.62/8.02}  & \textbf{10.1/9.55} & \textbf{12.19/11.15} & \textbf{20.25/19.01} \\
\bottomrule
\end{tabular}}
\begin{tablenotes}
    \item[\dag] 
    "10.23/9.59" represents the performance on NVIDIA RTX 3070 and Tesla A100 GPU, respectively.
     \item[\dag]
     “M1”: {\tt BD-NTT-2}, “M2”: {\tt OSSP-BD-NTT-2}, “M3”: {\tt OSSP-PCS-BD-NTT-2}
     \item
     “M4”: {\tt TA-NTT-2}, “M5”: {\tt OSSP-TA-NTT-2}, “M6”: {\tt OSSP-PCS-TA-NTT-2}
  \end{tablenotes}
  \end{threeparttable}
\end{center}
\vspace{-0.6cm}
\end{table}

\begin{figure*}[t]
\setlength{\abovecaptionskip}{0.1cm}
\centering
\includegraphics[width=\linewidth]{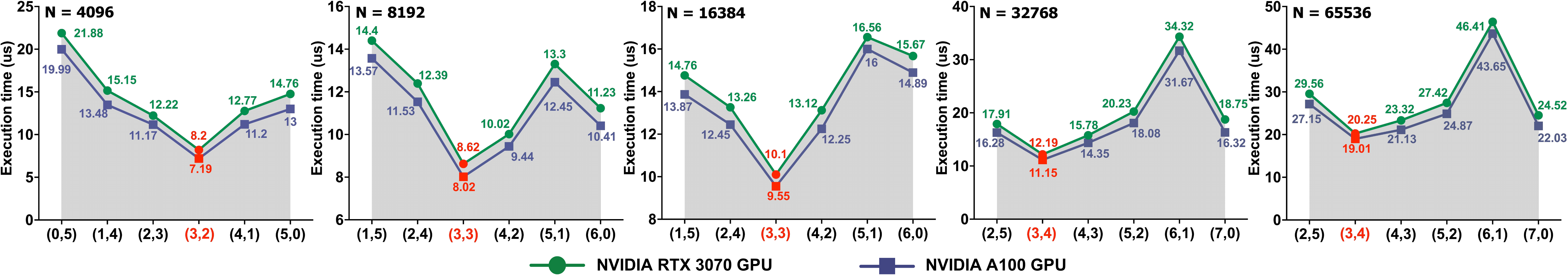}
\caption{Performance evaluation (microsecond) of the {optimal synchronization switching point-based} NTT ({\tt OSSP-PCS-TA-NTT-2}) {with synchronization combinations ($I_1$, $I_2$) of inter-block global synchronizations \textit{with} {\tt PCS} method (\textit{Phase 1}) and intra-block local synchronizations \textit{without} {\tt PCS} method (\textit{Phase 2}).}}
\label{fig:ntt-sensitivity}
\end{figure*}

\begin{table*}[t]
\begin{center}
\def\arraystretch{1.1}%
\setlength{\belowcaptionskip}{0cm} 
\setlength{\abovecaptionskip}{0.1cm} 
\caption{{Performance (millisecond) of CKKS Homomorphic Addition, Homomorphic Multiplication, Homomorphic Rotation and speedup \bm{$\mathcal S_1$} (SEAL vs. Chameleon), \bm{$\mathcal S_2$} (OpenFHE vs. Chameleon) and \bm{$\mathcal S_3$} (HElib vs. Chameleon).} }
% Note that in hybrid key switching mode, SEAL and Chameleon use $dnum=1$, whereas OpenFHE uses $dnum=2$.
\label{tab:ckks-performance}
% \begin{threeparttable}
\scriptsize
\centering
\setlength{\tabcolsep}{0.2mm}{
\begin{tabular}{cccccccccccccccccccccccc}
\toprule
\multirow{2}{*}{\textbf{log\textit{N}}}&
\multirow{2}{*}{\textbf{log\textit{Q}}}&
\multicolumn{7}{c}{\textbf{Homomorphic Addition}} & \multicolumn{7}{c}{\textbf{Homomorphic Multiplication}} & \multicolumn{7}{c}{\textbf{Homomorphic Rotation}} \\
 \cmidrule(r){3-9} \cmidrule(r){10-16} \cmidrule(r){17-23} 
  & & \textbf{SEAL} & \textbf{Open.} &  \textbf{HElib}    &  \textbf{Chameleon}  &  \bm{$\mathcal S_1$} &  \bm{$\mathcal S_2$} & \bm{$\mathcal S_3$} &  \textbf{SEAL}  & \textbf{Open.} &  \textbf{HElib}  &  \textbf{Chameleon} & \bm{$\mathcal S_1$}  &  \bm{$\mathcal S_2$} &  \bm{$\mathcal S_3$} &
  \textbf{SEAL}  & \textbf{Open.} &  \textbf{HElib}  &  \textbf{Chameleon} & \bm{$\mathcal S_1$}  &  \bm{$\mathcal S_2$} &  \bm{$\mathcal S_3$}  \\
\midrule
13      & 218	  	
        & 0.24   & 0.63   & 0.05   	    & 0.01	  & 24.0$\times$ & 63.0$\times$	& 5.0$\times$   	
        & 4.51   & 6.76   & 5.89	              & 0.11	 & 41$\times$ & 61.5$\times $ & 53.5$\times$ 
        & 3.69   & 7.03     & 5.42         	    & 0.1		& 36.9$\times$ & 70.3$\times$ & 54.2$\times$
        \\
% \midrule
14      & 438     		
        & 0.88   & 1.12   & 0.15         		 & 0.02		& 44.0$\times$ & 56.0$\times$ & 7.5$\times$   
        & 26.22  & 22.83  & 19.8              & 0.32		& 81.9$\times$ & 71.3$\times$ & 61.9$\times$
        & 22.51   & 23.92    & 18.71     & 0.29		& 77.6$\times$ & 82.5$\times$ & 64.5$\times$
        \\
% \midrule
15      & 881	  
        & 3.32   & 5.78    & 0.59	          	& 0.03      & 110.7$\times$ &192.7$\times$ & 19.7$\times$    
        & 149.57 & 98.76   & 94.32             & 1.29		& 115.9$\times$ & 76.6$\times$   & 73.1$\times$
        & 136.87   & 105.92      & 92.43	        & 1.22   & 112.2$\times$ & 86.8$\times$   & 75.8$\times$
        \\
\bottomrule
\end{tabular}}
\end{center}
\vspace{-0.5cm}
\end{table*}

\subsection{Performance Analysis of {Optimal Synchronization Switching Point-based} NTT}

\textbf{Performance benefits of optimal synchronization switching point method.}
Unlike previous hybrid implementations\cite{ozerkefficient}, our approach simplifies execution by requiring only a single kernel launch. 
This reduces the overhead of multiple launches and improves overall efficiency.
Table \ref{tab:ntt-mix} shows the performance of {optimal synchronization switching point-based} NTT. 
Using the \textit{Local Sync} synchronization method, both {\tt OSSP-BD-NTT-2} and {\tt OSSP-TA-NTT-2} achieve an average speedup of 1.19$\times$/1.21$\times$ and 1.17$\times$/1.16$\times$, respectively, over the {\tt BD-NTT-2} and {\tt TA-NTT-2}. 
Furthermore, by leveraging the polynomial coefficient shuffling method, {\tt OSSP-PCS-TA-NTT-2} delivers average performance gains of 1.15$\times$/1.15$\times$ and 1.16$\times$/1.16$\times$ over {\tt OSSP-BD-NTT-2} and {\tt OSSP-TA-NTT-2}. 
After applying all optimizations, {\tt OSSP-PCS-TA-NTT-2} achieves a maximum speedup of 1.48$\times$/1.58$\times$ compared to the {\tt BASE-NTT}.

\textbf{Sensitivity analysis of synchronization switching points.}  
Specifically, we list all possible combinations of $(I_1, I_2)$ for the two phases at different polynomial lengths. 
Here we take {\tt OSSP-PCS-TA-NTT-2} as an example. 
Figure \ref{fig:ntt-sensitivity} demonstrates performance at various switching points, highlighting the sensitivity of switching points {\tt SP} on different GPUs.
We observe similar performance trends on the RTX 3070 and A100 GPUs, with optimal performance when the number of data chunks is \textit{closest} to the number of SMs. 
The ideal switching point generates 64 data chunks for all polynomial lengths, matching the 46 and 108 SMs on the RTX 3070 and A100 GPUs, respectively.
Two scenarios should be noted: before the optimal switching point, the creation of 16 data chunks results in low hardware utilization; after it, 256 chunks are generated, exceeding the number of SMs. 
Although this increases utilization by scheduling multiple thread blocks per SM, it also adds synchronization overhead.
The previous design \cite{ozerkefficient, shivdikar2022accelerating} corresponds to combinations: (0,5) for length 4096, (1,5) for lengths 8192 and 16384, and (2,5) for lengths 32768 and 65536. 
For these polynomial lengths, our implementation delivers speedups of 2.67$\times$/2.78$\times$, 1.67$\times$/1.69$\times$, 1.46$\times$/1.45$\times$, 1.47$\times$/1.46$\times$, and 1.46$\times$/1.43$\times$, respectively.

\vspace{-0.15cm}
\subsection{Performance Evaluation of CKKS, TFHE and Scheme Switching}
\vspace{-0.15cm}

\textbf{Performance comparison with CKKS primitive implemented on CPUs.}
Building upon the {Chameleon}, we implement the CKKS scheme on a GPU and compare its performance against three widely used CPU-based libraries: {HElib}, {SEAL}, and {OpenFHE}. 
Notably, we extend {OpenFHE} benchmark to support evaluations for the CKKS scheme.
The performance results are shown in Table \ref{tab:ckks-performance}.

{\begin{itemize}[leftmargin=0.5cm, noitemsep, nolistsep]
    \item 
    \textbf{The speedup of {HEADD}.} 
    Chameleon achieves performance improvements of 110.7$\times$, 192.7$\times$, and 19.7$\times$ compared to SEAL, OpenFHE and HElib, respectively.

    \item 
    \textbf{The speedup of {HEMUL}.} 
    Our implementation achieves speedups of 115.9$\times$, 76.6$\times$, and 73.1$\times$ over SEAL, OpenFHE and HElib, respectively.
    
    \item 
    \textbf{The Speedup of {HEROT}.} 
    Chameleon delivers performance improvements of up to 112.2$\times$, 86.8$\times$, and 75.8$\times$ over HElib, SEAL, and OpenFHE, which offer performance comparable to HEMUL.
    
\end{itemize}}

\textbf{Performance comparison with CKKS primitive implemented on GPUs.}
Table\ref{tab:ckks-gpu-performance} list the performance of existing GPU-accelerated CKKS implementations. 
For consistency, we adopt the same parameter settings as {Over100x}, {TensorFHE}, and {HE-Booster}, specifically $(N, {\rm log}Q, L, dnum) = (2^{16}, 2305, 44, 45)$. 
In contrast, {Phantom} uses similar parameters of $(2^{16}, 1700, 41, 42)$. 
The results indicate that for homomorphic multiplication, Chameleon achieves performance gains of 3.12$\times$, 1.72$\times$, 1.58$\times$, and 1.23$\times$ over {Over100x}, {HE-Booster}, {Phantom}, and {TensorFHE}, respectively. 
Comparable improvements are also observed in homomorphic rotation. 
Note that {Over100x} uses performance data conducted on an NVIDIA Tesla V100 GPU. 
Besides, in terms of the amortized bootstrapping time, Chameleon delivers significant performance gains, achieving speedups of 2.11$\times$, 1.96$\times$, and 1.15$\times$ over {Over100x}, {HE-Booster}, and {TensorFHE}, respectively.

\begin{table}[h]
\begin{center}
\setlength{\belowcaptionskip}{0cm} 
\setlength{\abovecaptionskip}{0.1cm} 
\caption{Performance comparison of various CKKS implementations on GPUs.}
\label{tab:ckks-gpu-performance}
\begin{threeparttable}
\scriptsize
\centering
\def\arraystretch{1.1}%
\setlength{\tabcolsep}{0.7mm}{
\begin{tabular}{c|ccccc}
\toprule
\textbf{Operations}&\textbf{Over100x}&\textbf{HE-Booster}&\textbf{Phantom}& \textbf{ {TensorFHE}} &\textbf{Chameleon}\\
\hline
% \hline
\textbf{{HEMUL}} (ms)   & 17.4  	& 9.32 	&  8.56 &   6.65 	& \textbf{5.42} 			     \\
% \hline
Speedup & \textbf{3.21$\times$}  	& \textbf{1.72$\times$} 	&  \textbf{1.58$\times$} &   \textbf{1.23$\times$} 	& \textbf{1.0$\times$} 			 \\
% \hline
\textbf{{HEROT}} (ms) 	& 16.83 	& 8.87	&  8.49 &   6.66 	& \textbf{5.13}	    	 \\
% \hline
Speedup & \textbf{3.28$\times$}  & \textbf{1.67$\times$} 	&  \textbf{1.6$\times$}  &   \textbf{1.3$\times$} 	& \textbf{1.0$\times$} 			     \\
% \hline
\textbf{{Bootstrapping}} (ns) 	& 740 	& 685	&  --- &   404 	& \textbf{350}	    	 \\
Speedup & \textbf{2.11$\times$}  & \textbf{1.96$\times$} 	&  ---  &   \textbf{1.15$\times$} 	& \textbf{1.0$\times$} 			     \\
\hline
\textbf{GPU Type} & V100  & A100 	&  A100  &   A100 	& A100 			     \\
% \hline
\bottomrule
\end{tabular}}
\begin{tablenotes}
     \item[\dag] {Phantom} does not report the execution time for bootstrapping.
  \end{tablenotes}
  \end{threeparttable}
% \vspace{-0.5cm}
\end{center}
\end{table}

\begin{table}[t]
\def\arraystretch{1.1}%
\begin{center}
\setlength{\belowcaptionskip}{0cm} 
\setlength{\abovecaptionskip}{0.1cm} 
\caption{Performance (millisecond) and throughput (operations per second) for TFHE bootstrapping.}
\label{tab:tfhe-performance}
\begin{threeparttable}
\scriptsize
\centering
\setlength{\tabcolsep}{2.6mm}{
\begin{tabular}{c|ccc}
\toprule
\multirow{2}{*}{\textbf{Scheme}}&
 \multicolumn{3}{c}{{$(n_{\rm ckks}, n_{\rm lwe}, k, g_b)$}} \\
 \cmidrule(r){2-4} 
  &   \textbf{(1024,500,1,2)} & \textbf{(1024,630,1,3)} &  \textbf{(2048,592,1,3)}  \\
\midrule
 \textbf{TFHE-rs}         & 4.18/239/\textbf{2.77$\times$}     &  5.37/186/\textbf{3.22$\times$}            & 11.2/89/\textbf{4.87$\times$}     \\
\hline
 \textbf{cuFHE}           & 2.02/495/\textbf{1.34$\times$}     &  2.44/410/\textbf{1.47$\times$}            &   ---  	   \\
 \textbf{nuFHE}           & 2.03/492/\textbf{1.34$\times$}     &  2.45/408/\textbf{1.47$\times$}            &   ---  	   \\
 \textbf{CPU-GPU-TFHE}    & 2.09/478/\textbf{1.38$\times$}     &  2.53/395/\textbf{1.51$\times$}            &   --- 	   \\
% \midrule
 \textbf{Chameleon}    & 1.51/667/\textbf{1.0$\times$}     &  1.67/599/\textbf{1.0$\times$}            &  2.3/435/\textbf{1.0$\times$}   	   \\
\bottomrule
\end{tabular}
}
\begin{tablenotes}
     \item[\dag]
     "---" indicates that these implementations do not support the current parameter.
  \end{tablenotes}
  \end{threeparttable}
\end{center}
\vspace{-0.5cm}
\end{table}

\textbf{Performance comparison of TFHE bootstrapping on CPUs and GPUs.} 
Table \ref{tab:tfhe-performance} demonstrates the performance of TFHE gate bootstrapping, comparing the state-of-the-art CPU implementation, {TFHE-rs}, with three GPU-based ones: {cuFHE}, {nuFHE}, and {CPU-GPU-TFHE}. 
Using three parameter sets with security levels of 80, 110, and 128 bits, Chameleon achieves speedup of 3.22$\times$, 1.47$\times$, 1.47$\times$, and 1.51$\times$ over {TFHE-rs}, {cuFHE}, {nuFHE}, and {CPU-GPU-TFHE}, respectively.

\textbf{Performance of scheme switching between CKKS and TFHE on CPUs.}
Finally, we evaluate the performance of scheme switching between CKKS and TFHE, emphasizing a detailed breakdown of significant paths in this process, and compare our findings with the state-of-the-art CPU implementation, {Pegasus}~\cite{lu2021pegasus}.
The results show that Chameleon achieves an average speedup of 67.3$\times$ compared to Pegasus. 
Specifically, sample extract and repack boost performance by 106.9$\times$ and 123.3$\times$, respectively, due to effective homomorphic multiplication, rotation operations, and low-level NTT in these phases. 
In contrast, Slot2Coeff involves only a few homomorphic rotations and low-overhead plaintext-ciphertext multiplication, resulting in an speedup of 60.5$\times$. In addition, using a CMux-level parallelization approach for LUT evaluation leads to performance gains of 65.6$\times$, but the mandatory sequential execution flow from gate bootstrapping limits the acceleration potential.

\section{Related Work}

\textbf{GPU-based NTT acceleration. }
${\ddot {\mathrm O}}$zerk et al. \cite{ozerkefficient} propose an efficient NTT acceleration design using stage-by-stage synchronization. 
The single-kernel method loads the entire polynomial into one thread block, resulting in limited parallelism, whereas the multi-kernel design necessitates repeated kernel launches, leading to significant overhead.
The hybrid approach combines both methods to address these issues but lacks an analysis of the optimal switching point.
Other works \cite{shivdikar2022accelerating,zhai2022accelerating,yang2022cuxcmp} adopt similar acceleration strategies. 
In contrast, our implementation prioritizes reducing the number and scale of synchronization to boost performance. 
We also identify an optimal synchronization switching point in the hybrid approach, yielding up to 2.78$\times$ performance boosts. 
Besides, cuFFT \cite{cufft} offers an FFT implementation, and our results exhibit speedups of 2.07$\times$, 2.64$\times$, and 2.95$\times$ for the same polynomials, with reduced execution times of 17.86us vs. 8.62us, 26.7us vs. 10.1us, and 35.98us vs. 12.19us.

\textbf{GPU-based FHE acceleration.}
{Over100x}~\cite{jung2021over} introduces the first GPU-accelerated RNS-CKKS scheme with bootstrapping~\cite{cheon2018full,han2020better}, using intra-FHE and inter-FHE operation fusion, along with a critical operation reordering strategy, to address memory bandwidth bottlenecks.
{HE-Booster}~\cite{wang2023he} presents an efficient framework for accelerating polynomial arithmetic, supporting BGV and CKKS schemes. The framework focuses on five typical stages of polynomial arithmetic and includes optimization techniques, such as NTT with inter-thread local synchronization, to enhance performance.
{Phantom}~\cite{yang2024phantom} uses GPU to implement the BGV, BFV, and CKKS FHE schemes. It provides a unified framework integrating scheme functions and homomorphic operations, ensuring efficient and low-latency data access and transfer.
{TensorFHE}~\cite{fan2023tensorfhe} harnesses Tensor Core Units (TCUs) to accelerate NTT computations, with a focus on maximizing FHE operations to exploit data-level parallelism. 
{GME}~\cite{shivdikar2023gme} incorporates three micro-architectural extensions and compile-time optimizations for AMD CDNA GPUs. It evaluates CKKS performance using NaviSim, a cycle-level GPU simulator.
Focusing on the TFHE scheme \cite{chillotti2016faster}, two GPU-based FHE libraries, {cuFHE} \cite{cuFHE} and {nuFHE} \cite{NuFHE}, have been released. 
{CPU-GPU-TFHE} \cite{morshed2020cpu} presents new optimizations for additional speedup. 
It is worth noting that {Chameleon} is the first GPU-accelerated design to offer a comprehensive analysis and optimization targeted for scheme switching.

\begin{table}[t]
\def\arraystretch{1.1}%
\setlength{\tabcolsep}{1pt}
\begin{center}
\setlength{\belowcaptionskip}{0cm} 
\setlength{\abovecaptionskip}{0.1cm} 
\caption{Performance breakdown (millisecond) of Pegasus and Chameleon with different number of slots.}
\label{tab:sw}
\begin{threeparttable}
\scriptsize
\centering
\setlength{\tabcolsep}{1mm}{
\begin{tabular}{cc|cccccc}
\toprule
 \textbf{Scheme} & \textbf{$n_{\rm slot}$} & \textbf{Slot2Coeff} & \textbf{SamExt.} &
 \textbf{LUT} & \textbf{Repack} & \textbf{Total} & \textbf{Speedup} \\
\hline
 \multirow{3}{*}{\textbf{Pegasus}} 
 % & {4}   & 186  & 20  & 459  & 7550  & 8215   & 1.0$\times$ \\ 
 % & {16}  & 350  & 84  & 1848 & 8936  & 11218  & 1.0$\times$ \\ 
 & {64}  & 481  & 1077 & 28408 & 9938 & 39904  & \textbf{62.8$\times$} \\
 & {256}  & 946  & 4724 & 114731 & 34062 & 154463  & \textbf{70.1$\times$} \\
 & {1024}  & 1333  & 16675 & 447945 & 38232 & 504185  & \textbf{68.9$\times$} \\
\hline
 \multirow{3}{*}{\textbf{Chameleon}} 
 & {64} & 9 & 10 & 555 & 61 & 635 & 1.0$\times$ \\ 
 & {256} & 14 & 39 & 1979 & 173 & 2205  & 1.0$\times$\\ 
 & {1024} & 22 & 156 & 6826 & 310 & 7314 & 1.0$\times$\\
\bottomrule
\end{tabular}}
  \end{threeparttable}
\end{center}
\vspace{-0.8cm} %
\end{table}

\textbf{Hardware-based FHE acceleration.}
A substantial body of research is devoted to designing hardware accelerators on FPGA and ASIC platforms, aiming to further enhance computational efficiency.
On the one hand, {HEAX}~\cite{riazi2020heax} introduces a highly optimized architecture specifically designed for CKKS schemes. 
{FAB}~\cite{agrawal2022fab} presents a multi-FPGA FHE acceleration system to enhance performance, focusing on speeding up CKKS bootstrapping.
Existing schemes rely on fixed-size parameter sets~\cite{poppelmann2015accelerating,roy2019fpga, roy2018hepcloud}, which restrict the flexibility of FHE configurations across various application scenarios. Furthermore, these FPGA-based solutions also encounter significant productivity challenges.
On the other hand, {F1} \cite{samardzic2021f1}, {CraterLake} \cite{samardzic2022craterlake}, {BTS} \cite{kim2022bts}, {ARK} \cite{kim2022ark}, {BASALISC} \cite{geelen2022basalisc}, {SHARP}~\cite{kim2023sharp} and {BitPacker}~\cite{samardzic2024bitpacker} each propose ASIC accelerator architectures aimed at specific optimizations, such as programmability for flexible parameter settings, bootstrapping to support unbounded multiplicative depth, and efficient memory access for reduced data movement overhead. 
However, these designs have only been implemented and evaluated on simulators.

\section{Conclusion}
This paper presents Chameleon, an efficient GPU-based FHE acceleration framework that supports CKKS, TFHE, and scheme switching. 
Specifically, Chameleon features a scalable NTT acceleration design that adapts to polynomials of varying sizes, improving the efficiency of homomorphic operations in CKKS and TFHE. 
In addition, it is the first to optimize critical operations involved in scheme switching by introducing a CMux-level parallelization method for faster LUT evaluation and a homomorphic rotation-free matrix-vector multiplication to boost repack efficiency.
Experiments show that Chameleon outperforms state-of-the-art GPU implementations by 1.23$\times$ in CKKS homomorphic multiplication and 1.15$\times$ in bootstrapping. 
It also achieves speedups of 4.87$\times$ and 1.51$\times$ in TFHE gate bootstrapping compared to CPU and GPU versions, respectively. 
Finally, Chameleon produces 67.3$\times$ performance gain for scheme switching over CPU-based implementation.

\bibliographystyle{IEEEtran}
\bibliography{ref}

\begin{thebibliography}{10}
\providecommand{\url}[1]{#1}
\csname url@samestyle\endcsname
\providecommand{\newblock}{\relax}
\providecommand{\bibinfo}[2]{#2}
\providecommand{\BIBentrySTDinterwordspacing}{\spaceskip=0pt\relax}
\providecommand{\BIBentryALTinterwordstretchfactor}{4}
\providecommand{\BIBentryALTinterwordspacing}{\spaceskip=\fontdimen2\font plus
\BIBentryALTinterwordstretchfactor\fontdimen3\font minus \fontdimen4\font\relax}
\providecommand{\BIBforeignlanguage}[2]{{%
\expandafter\ifx\csname l@#1\endcsname\relax
\typeout{** WARNING: IEEEtran.bst: No hyphenation pattern has been}%
\typeout{** loaded for the language `#1'. Using the pattern for}%
\typeout{** the default language instead.}%
\else
\language=\csname l@#1\endcsname
\fi
#2}}
\providecommand{\BIBdecl}{\relax}
\BIBdecl

\bibitem{gentry2009fully}
C.~Gentry, ``Fully homomorphic encryption using ideal lattices,'' in \emph{Proceedings of the forty-first annual ACM symposium on Theory of computing}, 2009, pp. 169--178.

\bibitem{riazi2020heax}
M.~S. Riazi, K.~Laine, B.~Pelton, and W.~Dai, ``Heax: An architecture for computing on encrypted data,'' in \emph{Proceedings of the Twenty-Fifth International Conference on Architectural Support for Programming Languages and Operating Systems}, 2020, pp. 1295--1309.

\bibitem{agrawal2022fab}
R.~Agrawal, L.~de~Castro, G.~Yang, C.~Juvekar, R.~Yazicigil, A.~Chandrakasan, V.~Vaikuntanathan, and A.~Joshi, ``Fab: An fpga-based accelerator for bootstrappable fully homomorphic encryption,'' \emph{arXiv preprint arXiv:2207.11872}, 2022.

\bibitem{yang2023poseidon}
Y.~Yang, H.~Zhang, S.~Fan, H.~Lu, M.~Zhang, and X.~Li, ``Poseidon: Practical homomorphic encryption accelerator,'' in \emph{2023 IEEE International Symposium on High-Performance Computer Architecture (HPCA)}.\hskip 1em plus 0.5em minus 0.4em\relax IEEE, 2023, pp. 870--881.

\bibitem{samardzic2021f1}
N.~Samardzic, A.~Feldmann, A.~Krastev, S.~Devadas, R.~Dreslinski, C.~Peikert, and D.~Sanchez, ``F1: A fast and programmable accelerator for fully homomorphic encryption,'' in \emph{MICRO-54: 54th Annual IEEE/ACM International Symposium on Microarchitecture}, 2021, pp. 238--252.

\bibitem{kim2022bts}
S.~Kim, J.~Kim, M.~J. Kim, W.~Jung, J.~Kim, M.~Rhu, and J.~H. Ahn, ``Bts: An accelerator for bootstrappable fully homomorphic encryption,'' in \emph{Proceedings of the 49th Annual International Symposium on Computer Architecture}, 2022, pp. 711--725.

\bibitem{kim2023sharp}
J.~Kim, S.~Kim, J.~Choi, J.~Park, D.~Kim, and J.~H. Ahn, ``Sharp: A short-word hierarchical accelerator for robust and practical fully homomorphic encryption,'' in \emph{Proceedings of the 50th Annual International Symposium on Computer Architecture}, 2023, pp. 1--15.

\bibitem{wang2023he}
Z.~Wang, P.~Li, R.~Hou, Z.~Li, J.~Cao, X.~Wang, and D.~Meng, ``He-booster: An efficient polynomial arithmetic acceleration on gpus for fully homomorphic encryption,'' \emph{IEEE Transactions on Parallel and Distributed Systems}, 2023.

\bibitem{al2018high}
A.~Al~Badawi, B.~Veeravalli, C.~F. Mun, and K.~M.~M. Aung, ``High-performance fv somewhat homomorphic encryption on gpus: An implementation using cuda,'' \emph{IACR Transactions on Cryptographic Hardware and Embedded Systems}, pp. 70--95, 2018.

\bibitem{fan2023tensorfhe}
S.~Fan, Z.~Wang, W.~Xu, R.~Hou, D.~Meng, and M.~Zhang, ``Tensorfhe: Achieving practical computation on encrypted data using gpgpu,'' in \emph{2023 IEEE International Symposium on High-Performance Computer Architecture (HPCA)}.\hskip 1em plus 0.5em minus 0.4em\relax IEEE, 2023, pp. 922--934.

\bibitem{morshed2020cpu}
T.~Morshed, M.~M. Al~Aziz, and N.~Mohammed, ``Cpu and gpu accelerated fully homomorphic encryption,'' in \emph{2020 IEEE International Symposium on Hardware Oriented Security and Trust (HOST)}.\hskip 1em plus 0.5em minus 0.4em\relax IEEE, 2020, pp. 142--153.

\bibitem{lee2022low}
E.~Lee, J.-W. Lee, J.~Lee, Y.-S. Kim, Y.~Kim, J.-S. No, and W.~Choi, ``Low-complexity deep convolutional neural networks on fully homomorphic encryption using multiplexed parallel convolutions,'' in \emph{International Conference on Machine Learning}.\hskip 1em plus 0.5em minus 0.4em\relax PMLR, 2022, pp. 12\,403--12\,422.

\bibitem{cheon2023high}
J.~H. Cheon, M.~Kang, T.~Kim, J.~Jung, and Y.~Yeo, ``High-throughput deep convolutional neural networks on fully homomorphic encryption using channel-by-channel packing,'' \emph{Cryptology ePrint Archive}, 2023.

\bibitem{lee2023precise}
J.~Lee, E.~Lee, J.-W. Lee, Y.~Kim, Y.-S. Kim, and J.-S. No, ``Precise approximation of convolutional neural networks for homomorphically encrypted data,'' \emph{IEEE Access}, vol.~11, pp. 62\,062--62\,076, 2023.

\bibitem{bourse2018fast}
F.~Bourse, M.~Minelli, M.~Minihold, and P.~Paillier, ``Fast homomorphic evaluation of deep discretized neural networks,'' in \emph{Advances in Cryptology--CRYPTO 2018: 38th Annual International Cryptology Conference, Santa Barbara, CA, USA, August 19--23, 2018, Proceedings, Part III 38}.\hskip 1em plus 0.5em minus 0.4em\relax Springer, 2018, pp. 483--512.

\bibitem{lou2019she}
Q.~Lou and L.~Jiang, ``She: A fast and accurate deep neural network for encrypted data,'' \emph{Advances in neural information processing systems}, vol.~32, 2019.

\bibitem{boura2020chimera}
C.~Boura, N.~Gama, M.~Georgieva, and D.~Jetchev, ``Chimera: Combining ring-lwe-based fully homomorphic encryption schemes,'' \emph{Journal of Mathematical Cryptology}, vol.~14, no.~1, pp. 316--338, 2020.

\bibitem{lu2021pegasus}
W.-j. Lu, Z.~Huang, C.~Hong, Y.~Ma, and H.~Qu, ``Pegasus: bridging polynomial and non-polynomial evaluations in homomorphic encryption,'' in \emph{2021 IEEE Symposium on Security and Privacy (SP)}.\hskip 1em plus 0.5em minus 0.4em\relax IEEE, 2021, pp. 1057--1073.

\bibitem{chen2021efficient}
H.~Chen, W.~Dai, M.~Kim, and Y.~Song, ``Efficient homomorphic conversion between (ring) lwe ciphertexts,'' in \emph{International Conference on Applied Cryptography and Network Security}.\hskip 1em plus 0.5em minus 0.4em\relax Springer, 2021, pp. 460--479.

\bibitem{kim2023general}
A.~Kim, M.~Deryabin, J.~Eom, R.~Choi, Y.~Lee, W.~Ghang, and D.~Yoo, ``General bootstrapping approach for rlwe-based homomorphic encryption,'' \emph{IEEE Transactions on Computers}, 2023.

\bibitem{kim2022ark}
J.~Kim, G.~Lee, S.~Kim, G.~Sohn, J.~Kim, M.~Rhu, and J.~H. Ahn, ``Ark: Fully homomorphic encryption accelerator with runtime data generation and inter-operation key reuse,'' \emph{arXiv preprint arXiv:2205.00922}, 2022.

\bibitem{wang2023nttfusion}
Z.~Wang, P.~Li, R.~Hou, and D.~Meng, ``Nttfusion: Efficient number theoretic transform acceleration on gpus,'' in \emph{2023 IEEE 41st International Conference on Computer Design (ICCD)}.\hskip 1em plus 0.5em minus 0.4em\relax IEEE, 2023, pp. 357--365.

\bibitem{cheon2018bootstrapping}
J.~H. Cheon, K.~Han, A.~Kim, M.~Kim, and Y.~Song, ``Bootstrapping for approximate homomorphic encryption,'' in \emph{Advances in Cryptology--EUROCRYPT 2018: 37th Annual International Conference on the Theory and Applications of Cryptographic Techniques, Tel Aviv, Israel, April 29-May 3, 2018 Proceedings, Part I 37}.\hskip 1em plus 0.5em minus 0.4em\relax Springer, 2018, pp. 360--384.

\bibitem{roy2014compact}
S.~S. Roy, F.~Vercauteren, N.~Mentens, D.~D. Chen, and I.~Verbauwhede, ``Compact ring-lwe cryptoprocessor,'' in \emph{International workshop on cryptographic hardware and embedded systems}.\hskip 1em plus 0.5em minus 0.4em\relax Springer, 2014, pp. 371--391.

\bibitem{poppelmann2015high}
T.~P{\"o}ppelmann, T.~Oder, and T.~G{\"u}neysu, ``High-performance ideal lattice-based cryptography on 8-bit atxmega microcontrollers,'' in \emph{International conference on cryptology and information security in Latin America}.\hskip 1em plus 0.5em minus 0.4em\relax Springer, 2015, pp. 346--365.

\bibitem{sanders2010cuda}
J.~Sanders and E.~Kandrot, \emph{CUDA by example: an introduction to general-purpose GPU programming}.\hskip 1em plus 0.5em minus 0.4em\relax Addison-Wesley Professional, 2010.

\bibitem{jung2021over}
W.~Jung, S.~Kim, J.~H. Ahn, J.~H. Cheon, and Y.~Lee, ``Over 100x faster bootstrapping in fully homomorphic encryption through memory-centric optimization with gpus.'' \emph{IACR Cryptol. ePrint Arch.}, vol. 2021, p. 508, 2021.

\bibitem{9519408}
W.-j. Lu, Z.~Huang, C.~Hong, Y.~Ma, and H.~Qu, ``Pegasus: Bridging polynomial and non-polynomial evaluations in homomorphic encryption,'' in \emph{2021 IEEE Symposium on Security and Privacy (SP)}, 2021, pp. 1057--1073.

\bibitem{bae2023hermes}
Y.~Bae, J.~H. Cheon, J.~Kim, J.~H. Park, and D.~Stehl{\'e}, ``Hermes: efficient ring packing using mlwe ciphertexts and application to transciphering,'' in \emph{Annual International Cryptology Conference}.\hskip 1em plus 0.5em minus 0.4em\relax Springer, 2023, pp. 37--69.

\bibitem{liu2023efficient}
T.-L. Liu, Y.-T. Ku, M.-C. Ho, F.-H. Liu, M.-C. Chang, C.-F. Hsu, W.-C. Chen, and S.-H. Hung, ``An efficient ckks-fhew/tfhe hybrid encrypted inference framework,'' in \emph{European Symposium on Research in Computer Security}.\hskip 1em plus 0.5em minus 0.4em\relax Springer, 2023, pp. 535--551.

\bibitem{ozerkefficient}
O.~Ozerk, C.~Elgezen, A.~C. Mert, E.~Ozt{\"u}rk, and E.~Savas, ``Efficient number theoretic transform implementation on gpu for homomorphic encryption.''

\bibitem{xiao2010inter}
S.~Xiao and W.-c. Feng, ``Inter-block gpu communication via fast barrier synchronization,'' in \emph{2010 IEEE International Symposium on Parallel \& Distributed Processing (IPDPS)}.\hskip 1em plus 0.5em minus 0.4em\relax IEEE, 2010, pp. 1--12.

\bibitem{halevi2014algorithms}
S.~Halevi and V.~Shoup, ``Algorithms in helib,'' in \emph{Annual Cryptology Conference}.\hskip 1em plus 0.5em minus 0.4em\relax Springer, 2014, pp. 554--571.

\bibitem{sealcrypto}
``{M}icrosoft {SEAL} (release 4.1),'' \url{https://github.com/Microsoft/SEAL}, Jan. 2023, microsoft Research, Redmond, WA.

\bibitem{al2022openfhe}
A.~Al~Badawi, J.~Bates, F.~Bergamaschi, D.~B. Cousins, S.~Erabelli, N.~Genise, S.~Halevi, H.~Hunt, A.~Kim, Y.~Lee \emph{et~al.}, ``Openfhe: Open-source fully homomorphic encryption library,'' in \emph{proceedings of the 10th workshop on encrypted computing \& applied homomorphic cryptography}, 2022, pp. 53--63.

\bibitem{yang2024phantom}
H.~Yang, S.~Shen, W.~Dai, L.~Zhou, Z.~Liu, and Y.~Zhao, ``Phantom: a cuda-accelerated word-wise homomorphic encryption library,'' \emph{IEEE Transactions on Dependable and Secure Computing}, 2024.

\bibitem{TFHE-rs}
Zama, ``{TFHE-rs: A Pure Rust Implementation of the TFHE Scheme for Boolean and Integer Arithmetics Over Encrypted Data},'' 2022, \url{https://github.com/zama-ai/tfhe-rs}.

\bibitem{cuFHE}
W.~Dai, ``Cuda-accelerated fully homomorphic encryption library,'' \url{https://github.com/vernamlab/cuFHE}, 2018.

\bibitem{NuFHE}
B.~Opanchuk, ``Nufhe, a gpu-powered torus fhe implementation,'' \url{ https://github.com/nucypher/nufhe}, 2020.

\bibitem{shivdikar2022accelerating}
K.~Shivdikar, G.~Jonatan, E.~Mora, N.~Livesay, R.~Agrawal, A.~Joshi, J.~Abellan, J.~Kim, and D.~Kaeli, ``Accelerating polynomial multiplication for homomorphic encryption on gpus,'' \emph{arXiv preprint arXiv:2209.01290}, 2022.

\bibitem{zhai2022accelerating}
Y.~Zhai, M.~Ibrahim, Y.~Qiu, F.~Boemer, Z.~Chen, A.~Titov, and A.~Lyashevsky, ``Accelerating encrypted computing on intel gpus,'' in \emph{2022 IEEE International Parallel and Distributed Processing Symposium (IPDPS)}.\hskip 1em plus 0.5em minus 0.4em\relax IEEE, 2022, pp. 705--716.

\bibitem{yang2022cuxcmp}
H.~Yang, S.~Shen, Z.~Liu, and Y.~Zhao, ``cuxcmp: Cuda-accelerated private comparison based on homomorphic encryption,'' \emph{Cryptology ePrint Archive}, 2022.

\bibitem{cufft}
``cufft.'' \url{https://developer.nvidia.com/cufft}.

\bibitem{cheon2018full}
J.~H. Cheon, K.~Han, A.~Kim, M.~Kim, and Y.~Song, ``A full rns variant of approximate homomorphic encryption,'' in \emph{International Conference on Selected Areas in Cryptography}.\hskip 1em plus 0.5em minus 0.4em\relax Springer, 2018, pp. 347--368.

\bibitem{han2020better}
K.~Han and D.~Ki, ``Better bootstrapping for approximate homomorphic encryption,'' in \emph{Cryptographers’ Track at the RSA Conference}.\hskip 1em plus 0.5em minus 0.4em\relax Springer, 2020, pp. 364--390.

\bibitem{shivdikar2023gme}
K.~Shivdikar, Y.~Bao, R.~Agrawal, M.~Shen, G.~Jonatan, E.~Mora, A.~Ingare, N.~Livesay, J.~L. Abell{\'a}n, J.~Kim \emph{et~al.}, ``Gme: Gpu-based microarchitectural extensions to accelerate homomorphic encryption,'' in \emph{Proceedings of the 56th Annual IEEE/ACM International Symposium on Microarchitecture}, 2023, pp. 670--684.

\bibitem{chillotti2016faster}
I.~Chillotti, N.~Gama, M.~Georgieva, and M.~Izabachene, ``Faster fully homomorphic encryption: Bootstrapping in less than 0.1 seconds,'' in \emph{international conference on the theory and application of cryptology and information security}.\hskip 1em plus 0.5em minus 0.4em\relax Springer, 2016, pp. 3--33.

\bibitem{poppelmann2015accelerating}
T.~P{\"o}ppelmann, M.~Naehrig, A.~Putnam, and A.~Macias, ``Accelerating homomorphic evaluation on reconfigurable hardware,'' in \emph{International workshop on cryptographic hardware and embedded systems}.\hskip 1em plus 0.5em minus 0.4em\relax Springer, 2015, pp. 143--163.

\bibitem{roy2019fpga}
S.~S. Roy, F.~Turan, K.~Jarvinen, F.~Vercauteren, and I.~Verbauwhede, ``Fpga-based high-performance parallel architecture for homomorphic computing on encrypted data,'' in \emph{2019 IEEE International symposium on high performance computer architecture (HPCA)}.\hskip 1em plus 0.5em minus 0.4em\relax IEEE, 2019, pp. 387--398.

\bibitem{roy2018hepcloud}
S.~S. Roy, K.~J{\"a}rvinen, J.~Vliegen, F.~Vercauteren, and I.~Verbauwhede, ``Hepcloud: An fpga-based multicore processor for fv somewhat homomorphic function evaluation,'' \emph{IEEE Transactions on Computers}, vol.~67, no.~11, pp. 1637--1650, 2018.

\bibitem{samardzic2022craterlake}
N.~Samardzic, A.~Feldmann, A.~Krastev, N.~Manohar, N.~Genise, S.~Devadas, K.~Eldefrawy, C.~Peikert, and D.~Sanchez, ``Craterlake: a hardware accelerator for efficient unbounded computation on encrypted data.'' in \emph{ISCA}, 2022, pp. 173--187.

\bibitem{geelen2022basalisc}
R.~Geelen, M.~Van~Beirendonck, H.~V. Pereira, B.~Huffman, T.~McAuley, B.~Selfridge, D.~Wagner, G.~Dimou, I.~Verbauwhede, F.~Vercauteren \emph{et~al.}, ``Basalisc: Flexible asynchronous hardware accelerator for fully homomorphic encryption,'' \emph{arXiv preprint arXiv:2205.14017}, 2022.

\bibitem{samardzic2024bitpacker}
N.~Samardzic and D.~Sanchez, ``Bitpacker: Enabling high arithmetic efficiency in fully homomorphic encryption accelerators,'' in \emph{Proceedings of the 29th ACM International Conference on Architectural Support for Programming Languages and Operating Systems, Volume 2}, 2024, pp. 137--150.

\end{thebibliography}

\end{document}